\documentclass[preprint,aps,showpacs,preprintnumbers,amsmath,amssymb,superscriptaddress]{revtex4-1}

\usepackage{graphicx}
\usepackage{dcolumn}
\usepackage{multirow}
\usepackage{placeins}
\usepackage{xcolor}
\usepackage[title]{appendix}
\usepackage{soul}

\begin{document}

\title{Structural and electronic properties of the random alloy ZnSe$_x$S$_{1-x}$}

\date{\today}
      
\author{S. Sarkar}
\affiliation{Asia Pacific Center for Theoretical Physics, Pohang, 37673, Korea}

\author{O. Eriksson}
\affiliation{Department of Physics and Astronomy, Uppsala University, Box 516, SE-75120, Uppsala, Sweden}
\affiliation{School of Science and Technology, \"{O}rebro University, SE-70182 \"{O}rebro, Sweden}

\author{D.~D. Sarma}
\affiliation{Solid State and Structural Chemistry Unit, Indian Institute of Science, Bengaluru 560012, India}
\affiliation{CSIR-National Institute for Interdisciplinary Science and Technology (CSIR-NIIST), Industrial Estate P.O., Pappanamcode, Thiruvananthapuram 695019, India}

\author{I. Di Marco}
\email{igor.dimarco@apctp.org}
\email{igor.dimarco@physics.uu.se}
\affiliation{Asia Pacific Center for Theoretical Physics, Pohang, 37673, Korea}
\affiliation{Department of Physics and Astronomy, Uppsala University, Box 516, SE-75120, Uppsala, Sweden}
\affiliation{Department of Physics, POSTECH, Pohang, 37673, Korea}

\begin{abstract}  
In this article we employ density functional theory in the generalized gradient approximation to investigate the structural and electronic properties of the solid solution alloy $\text{Zn}\text{Se}_x\text{S}_{1-x}$ in the wurtzite structure. We analyzed the character of the bond lengths and angles at the atomic scale, using a supercell approach that does not impose any constraint on the crystal potential. 
We show that the bond lengths of pristine ZnS and ZnSe compounds are almost preserved between nearest neighbors, which is different from what would be anticipated if Vegard's law were valid at the atomic level. We also show that bond lengths start behaving in accordance to Vegard's law from the third shell of nearest neighbors onward, which in turn determines the average lattice parameters of the alloys determined by diffraction experiments. 
Fundamental building blocks around the anions are identified and are shown to be non-rigid but still volume preserving. 
Finally, the geometrical analysis is connected to the trend exhibited by the electronic structure, and in particular by the band gap. 
The latter is found to exhibit a small deviation from the linearity with respect to the Se concentration, in accordance to available experimental data.
By assuming a quadratic dependence, we can extract a bowing parameter and analyze various contributions to it with various calculations under selected constraints. The structural deformation in response to the doping process is shown to be the driving force behind the deviation from linearity. 
The difference in stiffness between ZnS and ZnSe is showed to play a key role in the asymmetric behavior of the bowing parameter observed in the S-rich and Se-rich regions.
\end{abstract}

\maketitle

\section{Introduction}
The mechanical, thermodynamic, electrical, magnetic  and optical properties of crystalline materials mainly depend on the crystal structure and chemical composition. Hence, material properties can be tuned by controlling either the structural degrees of freedom or the type of atoms entering in the solid~\cite{MLusi_Crygrowth_2018}. For example, interesting -- and often superior -- physical properties may occur in materials where compositional defects are purposely introduced~\cite{ZWu_CMS_2017,QZhang_ChemSci_2019}. Such defects may range from localized impurities or vacancies to the formation of a proper substitutional random alloy. The latter is of a particular interest in materials science, as its composition can be systematically adjusted between the pure components to explore how the physical properties evolve with it.
A change in composition is expected to induce structural changes, whose understanding is necessary to be able to drive the alloying process towards the desired materials properties.

One of the first observations of the relation between structure and composition was made by Vegard in 1921~\cite{LVegard_ZPhys_1921} and is currently known as Vegard's law~\cite{ARDenton_PRA_1991}. This law states that the lattice constant of an alloy can be expressed as the linear combination of the lattice constants of its components, weighted by their respective concentrations.
Vegard's law was introduced based on empirical considerations, but found a theoretical justification in the virtual crystal approximation (VCA), a decade later~\cite{LNord_AnnPhy_1931}. Let us consider an alloy of chemical formula $(\text{A}_{1-x},\text{B}_x)\text{C}$, where the random occurrence of two different atoms A and B at a given crystallographic site is possible. In VCA, one can describe this alloy via a fictitious system where atoms A and B are replaced by a virtual atom whose potential is assumed to be the compositional average of the potentials of A and B~\cite{JEBernard_PRB_1987,LBellaiche_PRB_2000}. 
As this virtual atom represents an interpolation of the behavior of the atoms in the parent compounds, VCA can easily describe the linear variation of many physical properties, including the lattice parameter as a function of composition, i.e. Vegard's law. 
Decades of work on alloys, however, have shown that many properties are not well described by VCA. At a fundamental level, VCA is a single-site effective medium theory, a feature that is shared by other well-known computational approaches to disorder as e.g. the coherent potential approximation (CPA)~\cite{faulkner1982modern}.
While these theories may potentially provide a good description of physical properties involving averages over multiple unit cells, their treatment becomes less and less adequate the more we restrict our domain of observation towards the atomic scale, where the individual A-C and B-C bond distances depend on the mutual interaction between the individual atoms at a very local level.
In this regard, extended X-ray absorption fine structure (EXAFS) experiments make it possible to measure the local bond distances directly. Several studies in the last three decades have demonstrated that the nearest neighbor bond lengths do not change drastically with composition, but remain close to their values in the pure systems~\cite{JAzoulay_PRB_1982,JCMikkelsen_PRL_1982,ABalzarotti_PRB_1984,ABalzarotti_PRB_1985,ZWu_PRB_1993,JPPorres_JAP_2004,SMukherjee_PRB_2014}. 
In two of the most recent studies~\cite{SMukherjee_PRB_2014,DD2}, we combined experimental measurements by EXAFS with theoretical data via density functional theory (DFT)~\cite{JonesRev,DeltaPaper} to identify unambiguously to what extent Vegard's law applies at the atomic scale in both cationic and anionic solid solutions, namely $\text{Zn}_x\text{Cd}_{1-x}\text{S}$ and $\text{Zn}\text{Se}_x\text{S}_{1-x}$. 
The usage of large supercells constructed via special quasi-random structures (SQS)~\cite{Zunger_PRL_1990} allowed us to avoid the limitations of the single-site effective medium theories, discussed above, providing a meaningful comparison between theory and experiment. 
These analyses have demonstrated that the individual bond lengths of the first and second nearest neighbors show a small to moderate change across the solid solution, while starting from the third coordination shell one can observe a behavior that is closer to what Vegard's law would predict. While cationic and anionic solid solutions show similar features, some distinctive differences are also present.  The most significant difference manifests primarily in the variations of the bond angles as a function of composition, which leads to identifying different kinds of elementary building blocks for the two types of solid solutions~\cite{SMukherjee_PRB_2014,DD2}. Understanding these differences is of paramount importance to shed light on the connection between microscopic changes at the atomic scale and macroscopic changes at the crystal scale, following Vegard's law.
In this sense, a limitation of the previous two studies~\cite{SMukherjee_PRB_2014,DD2} is that the EXAFS analysis yields only an average value for each bond distance. This is a gross approximation of the real world, where one expects a wide distribution of bond distances between two atoms, based on their local environment.
These features at the atomic scale are totally missed by EXAFS and hence were not addressed in our previous study, ignoring the rich information available in the distribution of bond distances. 
The present theoretical work deals with these distributions and not just an average number, providing microscopic insights impossible to obtain from any experiment that necessarily averages over a finite length scale, as in X-ray diffraction (XRD), or different local compositional distributions, as in EXAFS.
In addition, the cationic and anionic solid solutions we investigated were ordered in different crystal structures, respectively wurtzite and cubic zinc blende. These structures, being the experimental ones, are optimal for a quantitative comparison with EXAFS data, but they may complicate the comparison of the physical processes governing cationic substitution and anionic substitution. To remedy this ambiguity, in the present manuscript, we extend our theoretical calculations to the anionic solid solution $\text{Zn}\text{Se}_x\text{S}_{1-x}$ in the wurtzite structure. By comparing these results with those obtained for the zinc blende structure, as well as with those obtained for $\text{Zn}_x\text{Cd}_{1-x}\text{S}$, we provide a deeper analysis of the distribution of the bond lengths and bond angles. This allows us to clarify the nature of the building blocks dominating the doping process in binary alloys.

In addition to this structural analysis, we also employ our computational data to understand the evolution of the band gap in solid solution alloys. In this case, theory and experiment predict a quadratic behavior, whose shape is associated with a bowing parameter~\cite{JEBernard_PRB_1987}. 
Although this problem in $\text{Zn}\text{Se}_x\text{S}_{1-x}$ has already been addressed in previous theoretical works~\cite{JEBernard_PRB_1987,DMesri_CMS_2007,DLong_CMS_2018,Hussain2019MT}, we provide here a detailed and systematic analysis that takes advantage of the large computational capabilities offered by modern supercomputers.
In this way, we are able to understand the importance of long range correlation in the stabilization of the bowing parameter, as well as to elucidate the different mechanisms contributing to the changes of the band gap. The latter is shown to be qualitatively connected to a volumetric analysis of the local crystal structure, as well as to a difference in stiffness between ZnS and ZnSe. 

The present manuscript is organized as follows. After this Introduction constituting Section I, the Methodology used for our study is presented in Section II. Then, Section III is dedicated to Results and Discussion, which are presented separately for the different coordination shells under consideration and finally for the electronic properties. Section IV contains our Conclusions to this investigation. For the sake of completeness, we have included more details on the structural analysis in the Appendices and in the Supplemental Material (SM)~\cite{SM}. The latter contains most of the data for the zinc blende structure, while selected plots are included in the main text for sake of comparison with the wurtzite structure.

\section{Methodology}
The electronic structure problem was addressed by using a projected augmented wave (PAW)~\cite{PEBlochl_PRB_1994} implementation of density functional 
theory, within the Vienna {\itshape{ab-initio}} simulation package 
(VASP)~\cite{GKresse_PRB_1996,GKresse_CompMatSci_1996}. 
The generalized gradient 
approximation (GGA)~\cite{JPPerdew_PRL_1996} in the Perdew-Burke-Ernzerhof (PBE) parametrization~\cite{JPPerdew_PRL_1997} was used for the exchange-correlation functional, unless stated otherwise. In the analysis of the band gap, we performed a few additional calculations with the modified Becke-Johnson (mBJ) exchange-potential in combination with correlations from the local density approximation (LDA)~\cite{MBJ_1,MBJ_2}.
The two end-members, i.e. ZnS and ZnSe, were modeled using the experimental wurtzite structure (space group P63mc)~\cite{EHKisi_Acta_1989,MPKulakov_InorgMat_1989}. The three intermediate compositions, namely ZnS$_{0.75}$Se$_{0.25}$, ZnS$_{0.50}$Se$_{0.50}$ and ZnS$_{0.25}$Se$_{0.75}$, were modeled using SQS~\cite{Zunger_PRL_1990}.
These were in turn generated using the Alloy Theoretic Automated Toolkit (ATAT)~\cite{Walle_Calphad_2002,Walle_Phaequ_2002,Walle_Calphad_2009}, based on a method derived from Monte Carlo simulations~\cite{Walle_Calphad_2013}.
Supercells of different size were considered, including 64, 96 and 128 atoms. For the largest supercells, all the distributions of the bond lengths and bond angles discussed in the manuscript are well converged. Hence, our analysis is focused on the results obtained for the supercells of 128 atoms.
In those, the pair correlation function associated to the SQS is found to match that of a perfectly random alloy till the 7$^{th}$ coordination shell.  
A full structural optimization was performed over all degrees of freedom and calculations were considered converged when forces on individual atoms became smaller than 10$^{-3}$ eV/\r{A}.
The Brillouin Zone was sampled by a $\Gamma$-centered Monkhorst-Pack mesh of $14 \times 14 \times 7$ for pristine ZnS/Se unit cell with 4 atoms. The 128 atom supercells were constructed by a $4 \times 4 \times 2$ repetition of the unit cell and for that a $8 \times 8 \times 10$ \textbf{k}-mesh was used. 
Along with this, an energy cutoff of 550 eV was used for the kinetic energy of the plane waves included in the basis. 
For sake of completeness, the convergence of the full structural optimization with respect to the energy cutoff of the plane waves and the sampling of the ${\bf{k}}$-mesh was accurately checked.  
The optimized lattice parameters obtained for ZnS were found to be 3.848, 3.848, and 6.311 \r{A}, which are on average $\sim$ 0.65\% larger than the corresponding experimental values. This small discrepancy can be linked to the well-known overestimation of the equilibrium volume in GGA~\cite{PHaas_PRB_2009,MSharma_PRB_2019,THomann_SSS_2006}, and it does not affect the analysis of our results.

\section{Results and Discussion}
\begin{figure}[bht]
\includegraphics[scale=0.5]{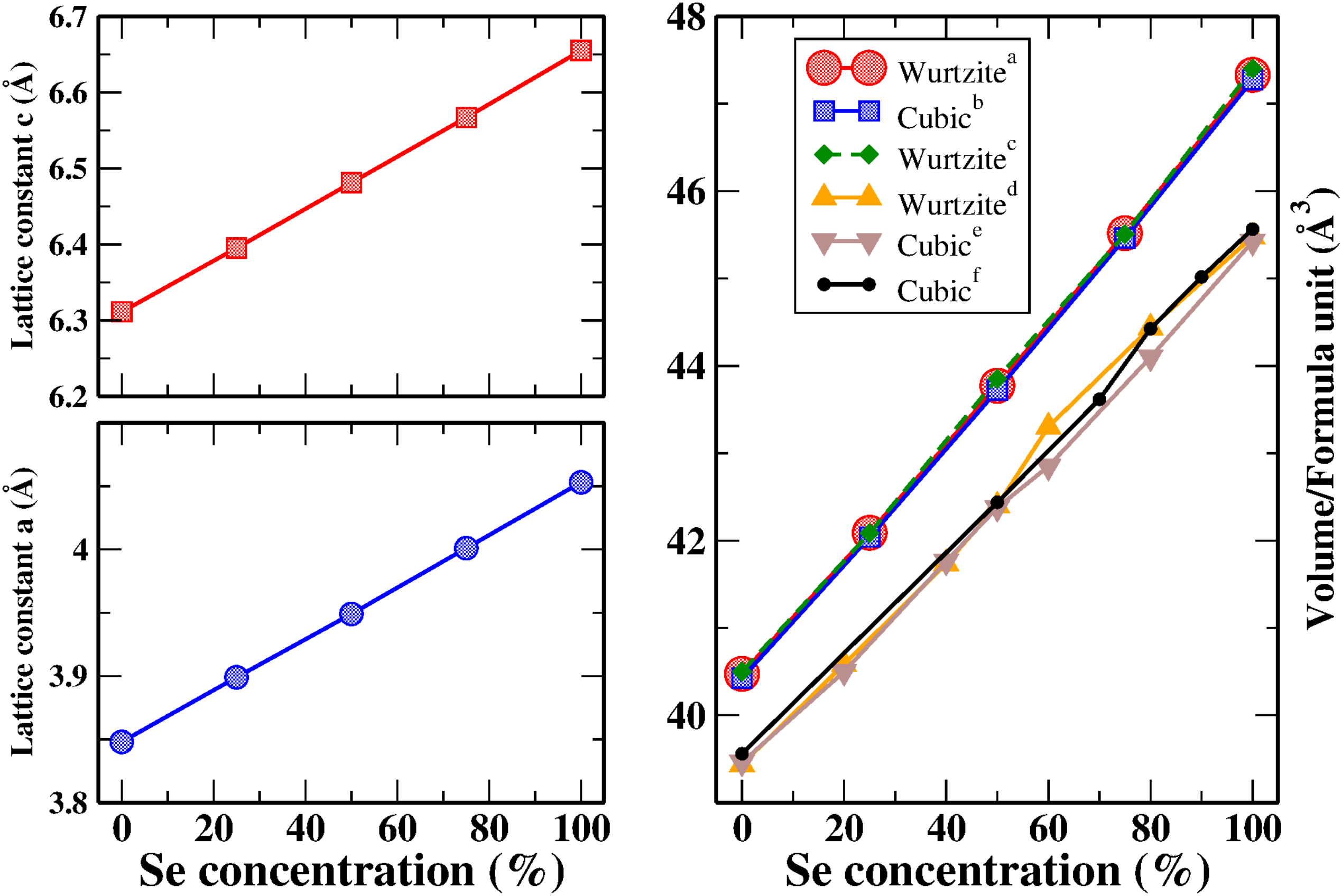}
\caption{Left panels: Variation of the lattice parameter $\bf{a = b}$ (bottom) and $\bf{c}$ (top) as a function of increasing Se concentration in $\text{Zn}\text{Se}_x\text{S}_{1-x}$ solid solutions. The linear variation of both is in agreement with the Vegard's law. Right panel: Variation of the volume per formula unit as a function of increasing Se concentration in wurtzite and cubic phases, as reported in different theoretical ($^a$this work; $^b$Ref.~\onlinecite{DD2}; $^c$Ref.~\onlinecite{Hussain2019MT}) and experimental ($^{d,e}$Ref.~\onlinecite{MAAviles_InoChem_2019}; $^f$Ref.~\onlinecite{DD2}) studies.}
\label{Fig1}
\end{figure}

Before addressing the behavior at the atomic scale of various shells of nearest neighbors, it is useful to look at the optimized values of the lattice parameters, which are directly connected to the macroscopic properties of the sample. The lattice parameters of the wurtzite structure, $a=b$ and $c$, as a function of increasing Se concentration are reported in the left panels of Fig.~\ref{Fig1}. As expected by Vegard's law, we can observe a linear increase when going from ZnS to ZnSe, which is also in line with what was reported by both experiment and theory~\cite{DD2}. Naturally, for the cubic phase we have only one lattice parameter, whose value is intermediate to the values of $a$ and $c$ found in the wurtzite structure.
The $c/a$ ratio can be inferred from the linear behavior of $a$ and $c$, and exhibits a perfectly linear behavior as well, going from 1.640 to 1.642.
For a comparison across different phases and with existing literature, it is convenient to plot the volume per formula unit for varying concentrations, reported in the right panel of Fig.~\ref{Fig1}. The present data (red circles) are shown to overlap almost perfectly with the data for the cubic phase (blue squares) from Ref.~\onlinecite{DD2}, as well as with those (green diamonds) obtained in a very recent theoretical study~\cite{Hussain2019MT}, despite much smaller supercells used there. 
This reflects what we pointed out in our recent work on the cubic phase, i.e.  the behavior of the lattice parameter, being  determined by a long-range  averaging over many unit cells, does not depend crucially on the precise description of the local environments in $\text{Zn}\text{Se}_x\text{S}_{1-x}$~\cite{DD2}.
The comparison with recently reported~\cite{DD2,MAAviles_InoChem_2019} experimental data (black dots, up and down triangles) is also favorable. Alongside the shift of the curves, associated to the overestimation of the bond-length mentioned above, we also observe that the experimental data show some irregularities in the Se-rich region. These can be attributed to the difficulties in separating cubic and wurtzite phases, which are mixed in the experiment, depending on the concentration~\cite{MAAviles_InoChem_2019}.

\subsection{First shell environment}

\begin{figure}[hbt]
\includegraphics[scale=0.5]{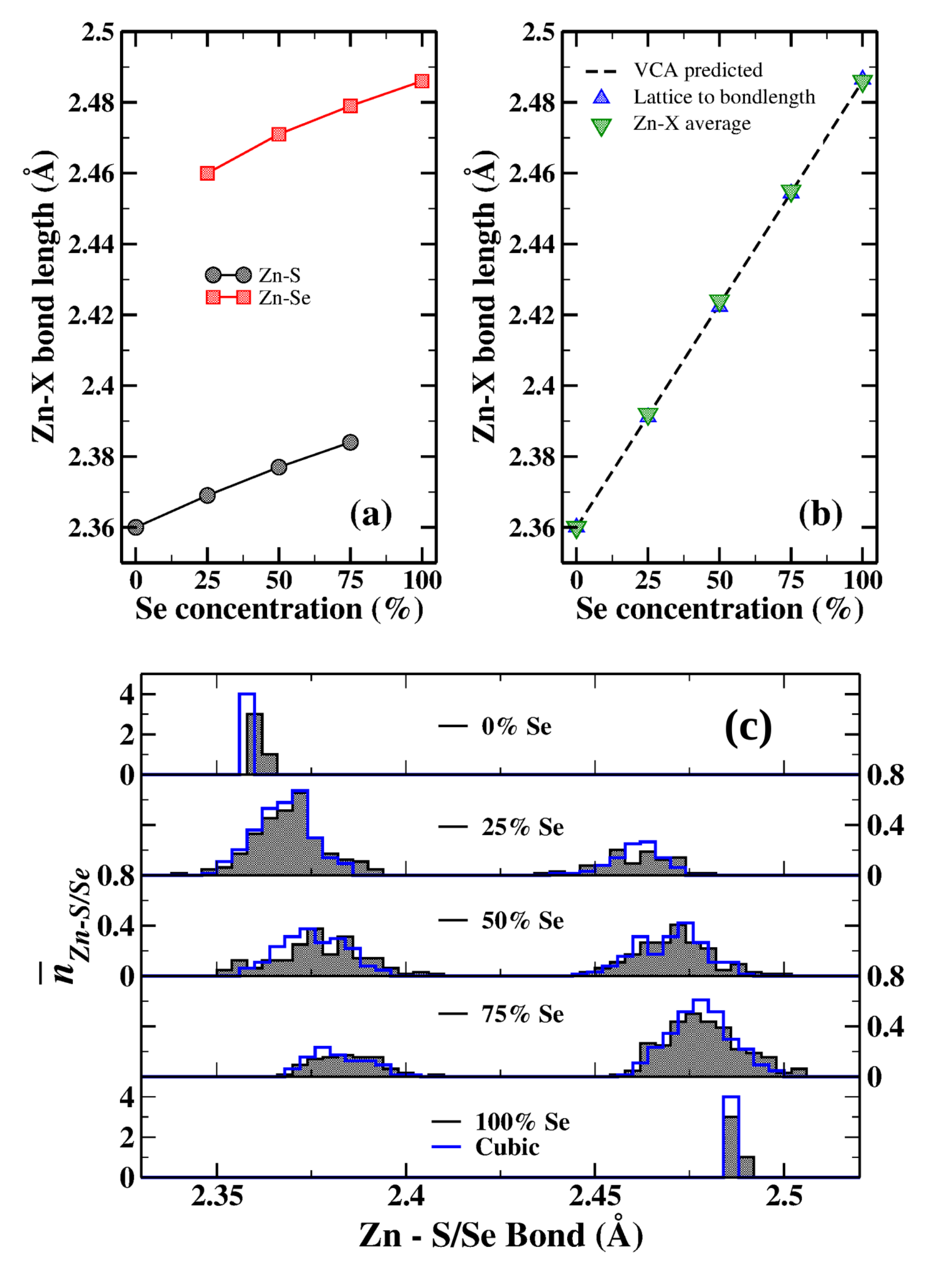}
\caption{(a) Variation in the Zn-S and Zn-Se nearest neighbor bond length  as a function of increasing Se concentration in $\text{Zn}\text{Se}_{x}\text{S}_{1-x}$ solid solutions.
(b) Average nearest neighbor Zn-$X$ ($X$ = S or Se) bond length (down triangles) compared to a hypothetical VCA-like average Zn-$X$ bond length (dashed line) obtained connecting the Zn-$X$ bond lengths of the pure end members; the Zn-S bond length (up triangles) calculated by imposing optimized lattice parameters in the ZnS wurtzite unit cell is also shown. (c) Distribution of the nearest neighbor Zn-S/Se bonds in the solid solution alloy for increasing Se concentration (top to bottom), normalized with respect to the number of bonds per Zn atom (4); the unfilled blue lines represent data for the zinc blende structure. The details of the ranges and averages for each composition in the wurtzite phase are also tabulated in the SM~\cite{SM}.}
\label{Fig2}
\end{figure}

Next, we investigate the Zn-S and Zn-Se nearest-neighbor bond lengths for the first coordination shell. The variation in the average values of the nearest neighbor Zn-S (circles) and Zn-Se distances (squares) as a function of increasing Se concentration is shown in Fig.~\ref{Fig2}(a). The presence of two distinct distributions of bond lengths represent a strong deviation from simplistic expectations based on Vegard's law. The latter should basically correspond to results obtained in VCA, which are reported in Fig.~\ref{Fig2}(b) as a black dashed curve. They show a perfect agreement with those obtained by converting the lattice parameters in effective average bond lengths (up triangles). This is done by imposing the optimized lattice parameters from the alloy systems in the ZnS wurtzite unit cell and then measuring the Zn-S bond lengths.  Interestingly, this behavior would also be recovered by taking a full average of the Zn-$X$ ($X$ = S or Se) bond lengths, which will be analyzed below. Overall, Fig.~\ref{Fig2}(a) and Fig.~\ref{Fig2}(b) demonstrate that the individual Zn-S/Se bonds try to retain values close to those obtained in the pure end-members, without following Vegard's law, which would instead dictate a much larger change. 

\FloatBarrier
\FloatBarrier

More details on the distribution of the nearest neighbor bond lengths are obtained by examining the corresponding histograms, reported in Fig.~\ref{Fig2}(c). This picture shows clearly the presence of two distinct distributions, each with a spread of $\sim$ 0.05 {\AA}  around the average Zn-S/Se values. This spread is not small with respect to the bond length but is still much smaller than the difference between the average Zn-S/Se values. For reference, we provide a more compact overview of the data shown in these histograms, as well as in the following ones, in a series of Tables included in the SM~\cite{SM}.

 The analysis of the previous data show that the average value of all the Zn-S and Zn-Se bonds present in the system together closely follows Vegard's law (or equivalently what would have been predicted by the VCA). In addition, from Fig.~\ref{Fig2}(a), we can see that, though the average values of the Zn-S and Zn-Se bond lengths separately do not follow Vegard's law, both of them slightly increase as a function of increasing Se concentration. Finally, the histograms in Fig.~\ref{Fig2}(c) show that there is a range of values about the average Zn-S and Zn-Se values. All these observations can be understood in terms of two extreme situations where we substitute a single Se atom in an S-rich environment and vice versa. The presence of a larger Se atom in an S-rich region creates a tensile strain locally due to a local chemical pressure effect. This tensile strain increases the distance between the surrounding atoms and as a result the Zn-S bonds about the substituted Se atom increase from their natural values. This effect increases the average value of the Zn-S bond distances in the system slightly. So we expect the average Zn-S bond length to increase gradually with an increasing Se concentration as shown in Fig.~\ref{Fig2}(a). Exactly the opposite mechanism takes place when we substitute a smaller S atom in a Se-rich region and hence, the average Zn-Se bond length gradually decreases with increasing S concentration. Now, in a realistic alloy system for any composition, there will not be  just one type of environment but a fluctuation in the local composition. This fluctuation in the local composition will give rise to the dispersion in the extents of the volume expansion or contraction, leading to a distribution of Zn-S and Zn-Se bond distances about the average values as we see in Fig.~\ref{Fig2}(c).
 These features are very similar to those found for the cubic zinc blende phase as shown by the blue lines in Fig.~\ref{Fig2}(c). The average values of the Zn-S/Se bonds in the two phases match perfectly, which is consistent with our previous interpretation in terms of local environment only. A minor difference is due to the spread of the distributions, which is $\sim$ 0.04 \r{A} in the cubic phase, i.e. about 0.01 \r{A} smaller than in the wurtzite phase. This is expected, as the wurtzite phase possesses a lower symmetry than the cubic phase, which makes one of the four bonds in the tetrahedral cage slightly bigger than the other three. This is clearly visible in the histograms for pristine ZnS and ZnSe and results into a wider distribution of the bond lengths at intermediate compositions. This diminished symmetry also introduces an ambiguity in how the atoms are included in different shells of neighbors. For reference, our precise partitioning is reported in Table~\ref{TableA1}, for both wurtzite and zinc blende phases.

 \subsection{Second shell environment} 

\begin{figure}[hb]
\includegraphics[scale=0.5]{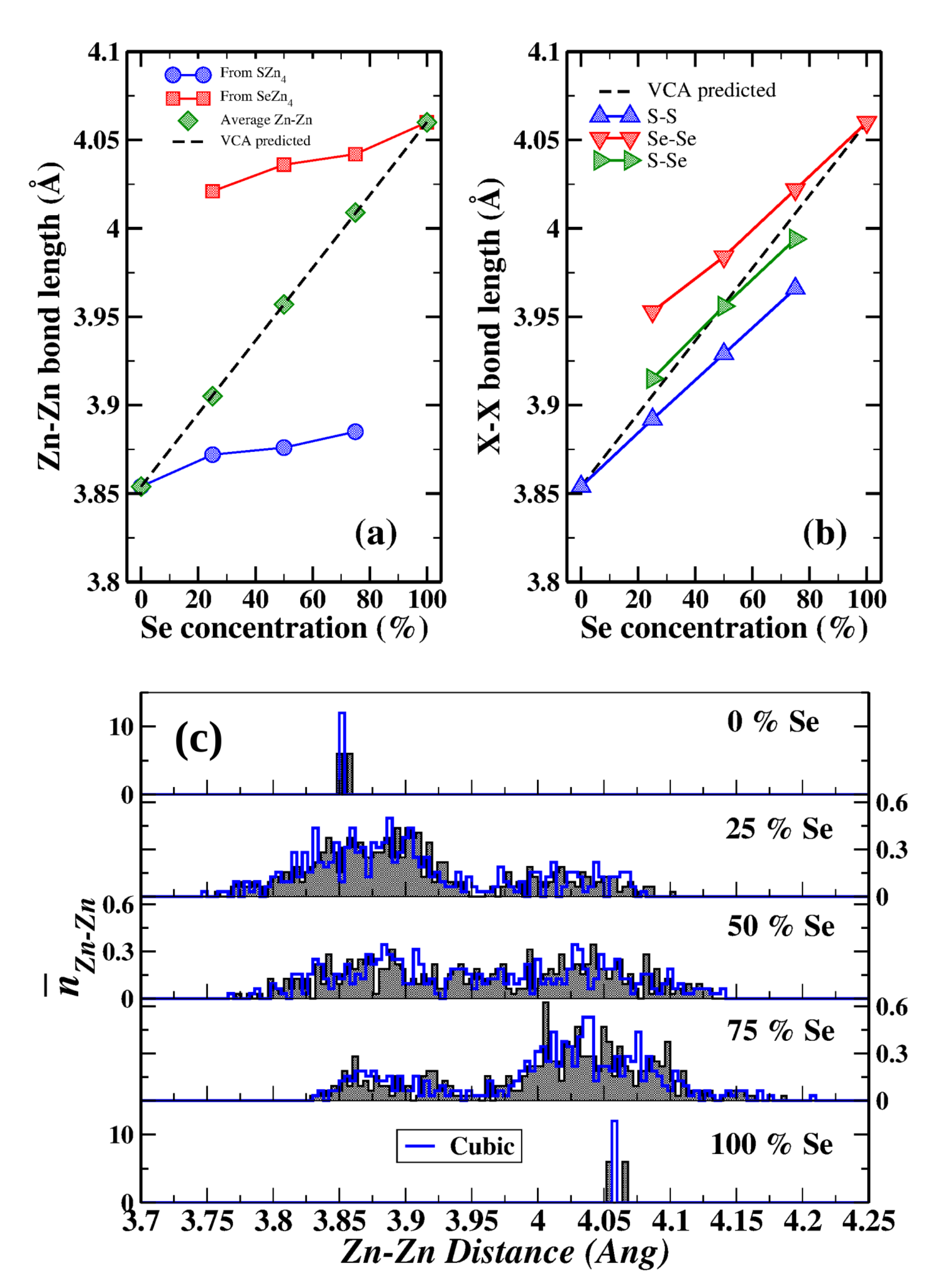}
\caption{Variation in the (a) Zn-Zn second neighbor bond length and (b) $X-X$ ($X$ = S or Se) second neighbor bond length  as a function of increasing Se concentration in $\text{ZnSe}_{x}\text{S}_{1-x}$. 
(c) Distribution of the nearest neighbor Zn-Zn bonds in the solid solution alloy for increasing Se concentration (top to bottom), normalized with respect to the number of bonds per Zn atom (12); the unfilled blue lines represent data for the zinc blende structure. For the alloy systems we see a clear bimodal distribution corresponding to the Zn-Zn bonds coming from ZnS and ZnSe type clusters respectively. More quantitative details on the range and averages of the distributions are provided in the SM~\cite{SM}.}
\label{Fig3}
\end{figure}

In the ideal wurtzite structure, the second coordination shell of any cation (anion) consists of 12 cations (anions). The only difference between the cationic and anionic sublattices in the case of the alloy systems under study is that, for the cationic sublattice, 12 atoms around each Zn atom are Zn atoms, but for the anionic sublattice both S and Se atoms are present around any S or Se atom.  
The distribution of the calculated Zn-Zn second neighbor bonds in the cationic sublattice is illustrated in Fig.~\ref{Fig3}(c). For all the intermediate concentrations, the distributions are clearly bimodal, albeit the peaks are less marked than those associated to the two separate nearest-neighbor distributions plotted in Fig.~\ref{Fig2}(c). This bimodality indicates the presence of a group of shorter and larger Zn-Zn bonds coming from the  SZn$_4$ and SeZn$_4$ tetrahedral units present in the system. Furthermore, it allows us to calculate separate average values for the Zn-Zn distance coming from SZn$_4$ and SeZn$_4$ tetrahedral units, shown as respectively circles and squares in Fig.~\ref{Fig3}(a). In the same plot, we also report the composition weighted average of both type of Zn-Zn bonds (diamond), as a function of increasing Se concentration. The black dotted line represents the variation of a single hypothetical  Zn-Zn bond that would have been predicted by VCA. We can see that the shorter and larger Zn-Zn bonds separately do not follow Vegard's law, which suggests that the Zn-Zn distances around S/Se atoms in the alloy systems retain values close to their  values in the pure systems. However, as for the first shell, the average value of all the second neighbor Zn-Zn bond lengths lies very close to the black dotted line corresponding to the VCA results. 

\FloatBarrier
\FloatBarrier

\begin{figure}[]
    \includegraphics[scale=0.6]{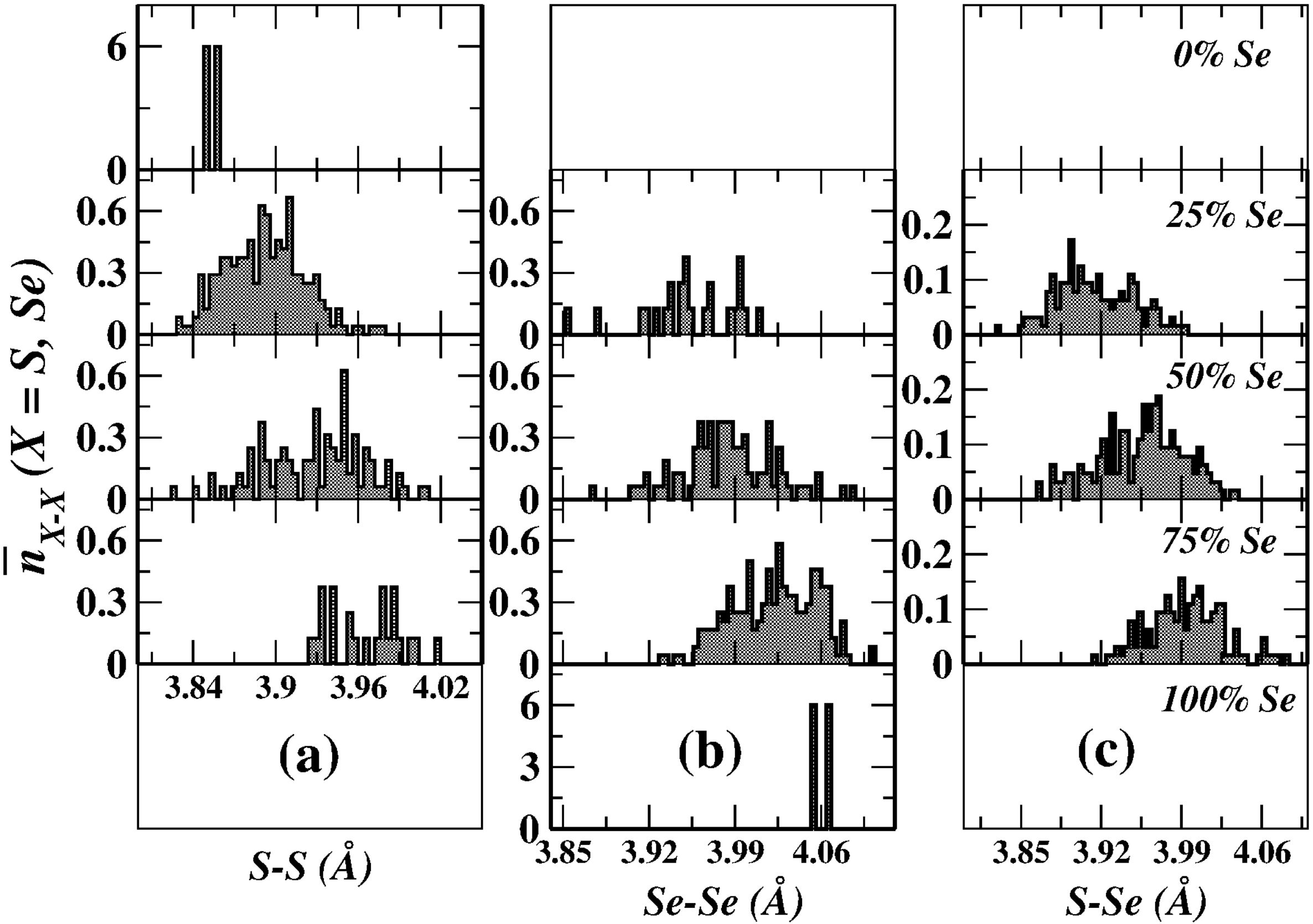}
  \caption{Distribution of the (a) S-S, (b) Se-Se, and (c) S-Se second neighbor bonds in the solid solution alloy for increasing Se concentration (top to bottom), normalized with respect to the number of bonds per S/Se atom (12), weighted by their relative concentration. More quantitative details on the range and averages of the distributions are provided in the SM~\cite{SM}.} 
  \label{Fig4}  
\end{figure}

We can then proceed to the analysis of the S-S, Se-Se, and S-Se second neighbor distances around the Zn atoms. The corresponding histograms, reported in Figs.~\ref{Fig4}(a),~\ref{Fig4}(b) and~\ref{Fig4}(c), are very different from the other plots we showed so far. For all the intermediate concentrations, the distributions of the bond lengths show a significant overlap, which is in sharp contrast with the two well marked peaks visible in Fig.~\ref{Fig2}(c). Nevertheless, since these distributions describe different pairs of atoms, it is still possible to calculate meaningful average values. Fig.~\ref{Fig3}(b) shows the average values of the S-S (up triangle), Se-Se (down triangle), and S-Se (right triangle) second neighbor bonds as a function of increasing Se concentration. 
The black dotted line again represents the variation of a single hypothetical average $X-X$ ($X$ = S or Se) bond that would have been predicted by VCA. From Fig.~\ref{Fig3}(b) we can see that the second neighbor bond lengths from the anionic sublattice follow Vegard's law more closely compared to the second neighbor Zn-Zn - see Fig.~\ref{Fig3}(a) - or nearest neighbor Zn-S/Se bond lengths - see Fig.~\ref{Fig2}(a). 
This same trend can also be seen in the cubic case, where the average values as well as the range of the quantities described above closely match with the values obtained for the wurtzite system. For example, the distributions of the Zn-Zn bonds from the cubic phase are shown alongside that of the wurtzite phase in Fig.~\ref{Fig3}(c), with unfilled blue lines. We see a very good qualitative match between these two phases. The only minor difference between wurtzite and cubic structures is that the second shell of wurtzite ZnS and ZnSe consist of two different sets of bond with a difference of $\sim$ 0.01 \r{A}, whereas in the cubic structure the 12 bonds are all of the same length. This is also visible from the histograms of the two end members in Fig.~\ref{Fig2}(c).  As discussed above, these features are a result of the lower symmetry of the wurtzite phase, when compared to the zinc blende phase.

\FloatBarrier

\FloatBarrier

\begin{figure}[hbt]
\includegraphics[scale=0.5]{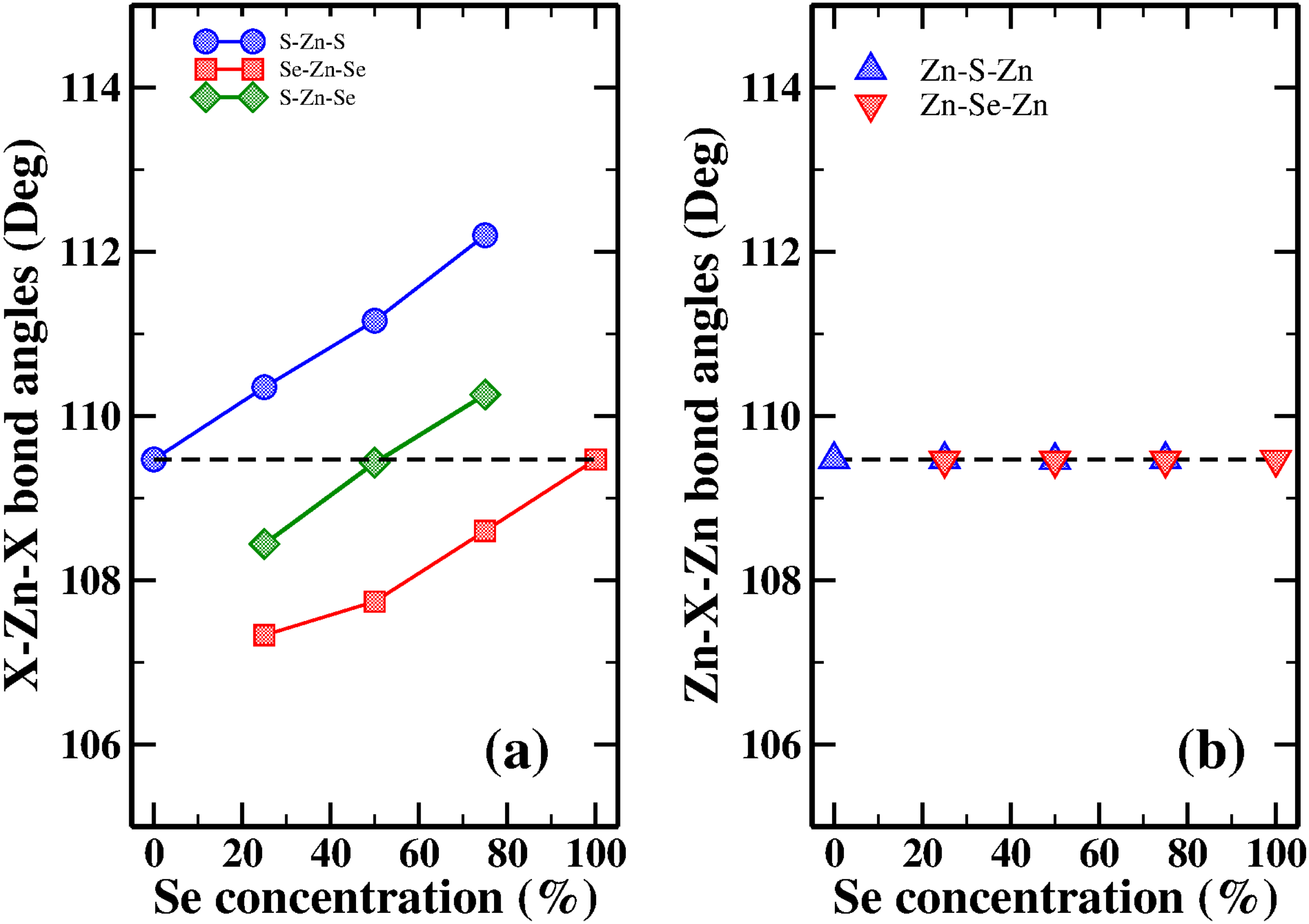}
\caption{Variation in the (a) $X$-Zn-$X$, and (b) Zn-$X$-Zn (X = S or Se) average bond angles as a function of increasing Se concentration in the $\text{ZnSe}_{x}\text{S}_{1-x}$ solid solutions.}
\label{Fig5}
\end{figure}

These observations in the solid solution $\text{ZnSe}_{x}\text{S}_{1-x}$ can only be understood by taking into account the variations in the cation centered $X$-Zn-$X$ and anion centered Zn-$X$-Zn ($X$ = S or Se) bond angles from their ideal value, obtained in pure ZnS or ZnSe. 
The ideal value here refers to the value of 109.47$^o$ that is obtained for a regular tetrahedron, where all four faces are equilateral triangles.
We first consider the cation-centered features.  The variations of the average values of the S-Zn-S (circle), Se-Zn-Se (square), and S-Zn-Se (diamond) bond angles as a function of increasing Se concentration are shown in Fig.~\ref{Fig5}(a). 
For completeness, the histograms of their distributions are reported in Appendix B, for all compositions.
The black dotted line in Fig.~\ref{Fig5}(a) corresponds to the ideal value of 109.47$^o$ across the series. From this figure, we can see that the S-Zn-S angle increases gradually with increasing Se concentration from its value in ZnS. On the contrary, the Se-Zn-Se angle decreases with increasing S concentration. The S-Zn-Se bond angle values lie in the region between the S-Zn-S and Se-Zn-Se values. Next, we proceed to analyze the two anion-centered bond angles, whose distributions are also detailed in Appendix B. In Fig.~\ref{Fig5}(b), we report the variations of the average values of the Zn-S-Zn and Zn-Se-Zn bond angles as a function of increasing Se concentration. We can see that the average values of both angles remain almost unchanged with respect to a change of composition. These trends are also similar to those reported for the cubic phase and, as for the bond lengths, the only difference concerns the spread of the distribution. The cation-centered angles in the cubic system show a larger spread of $\sim$ 1$^o$ compared to the wurtzite system.  For the anion-centered angles, it is the opposite. Here the spread in the wurtzite system is $\sim$ 1$^o$ larger compared to the cubic system.

Before providing a microscopic explanation of these results, it is useful to make a comparison of the present anionic solid solution with the cationic solid solution $\text{Zn}_x\text{Cd}_{1-x}\text{S}$ of Ref.~\cite{SMukherjee_PRB_2014}. While the nearest neighbor Zn-S/Cd-S bond behave similarly to Zn-S/Zn-Se bonds when concentration is changed, the behavior of the second shell environment changes with respect to the different doping process, affecting the cationic sublattice instead of the anionic one.
For $\text{Zn}_x\text{Cd}_{1-x}\text{S}$, the second coordination shell around any cationic site consists of both Zn and Cd atoms whereas any anionic site has only S atoms in its second coordination shell. This is just the opposite in the anion substituted $\text{ZnSe}_{x}\text{S}_{1-x}$ alloy system. As a result, in Ref.~\cite{SMukherjee_PRB_2014} we see that the second neighbor Zn-Zn, Cd-Cd, and Zn-Cd bonds in the cationic sublattice follows Vegard's law closely, similarly to what we report here for the S-S, Se-Se, and S-Se bonds in the anionic sublattice in $\text{ZnSe}_{x}\text{S}_{1-x}$. 
Concerning the bond angles, instead, we see that the average values of the cation centered S-Zn-S and S-Cd-S angles remain almost unchanged with respect to compositional change. It is the anion-centered bond angles, namely Zn-S-Zn, Cd-S-Cd, and Zn-S-Cd, that vary.
With increasing Cd concentration, the average Zn-S-Zn angle gradually increases from its ideal value, whereas the average Cd-S-Cd angle decreases from its ideal value with increasing Zn concentration. The Zn-S-Cd angle lies in between Zn-S-Zn and Cd-S-Cd and behaves similar to the S-Zn-Se angle in the anion substituted system.
This difference in the behavior of the second neighbor bonds between cation and anion substituted systems indicates that the structural changes at the atomic scale occurring in these two types of systems are very different, although the nearest neighbor cation-anion bonds follow the same trends.

\subsection{Third and fourth shell environment}
\begin{figure}[h]
\includegraphics[scale=0.5]{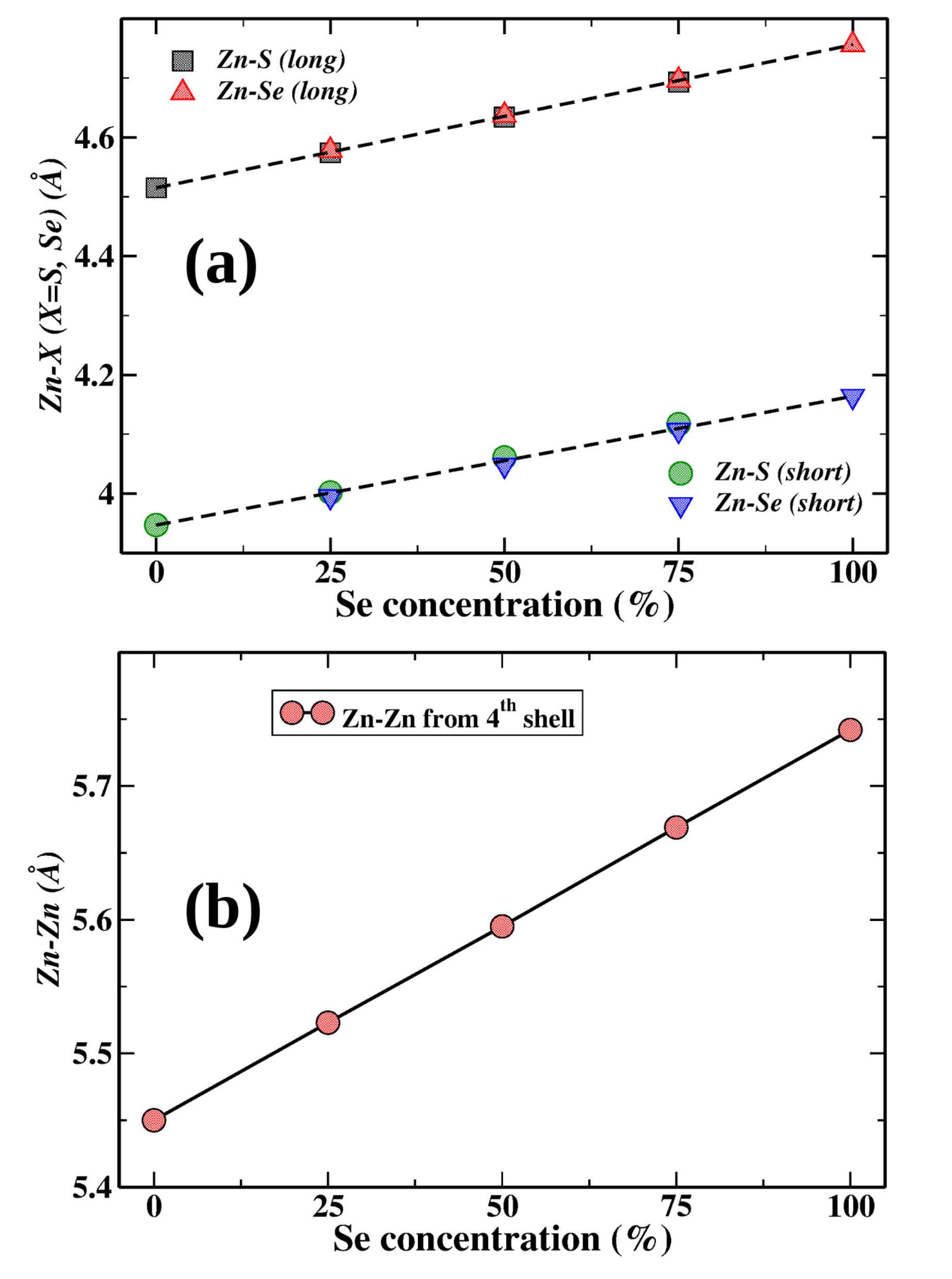}
\caption{(a) Variation in the Zn-$X$ ($X$ = S or Se) bond distance between Zn and S/Se atoms from the third coordination shell around Zn atoms, as a function of increasing Se concentration. (b) Variation in the Zn-Zn bond distance corresponding to the fourth coordination shell as a function of increasing Se concentration. The details of the average and range for these Zn-S, Zn-Se, and Zn-Zn bonds are given in the SM~\cite{SM}.}
\label{Fig6}
\end{figure}

\begin{figure}[]

    \includegraphics[scale=0.5]{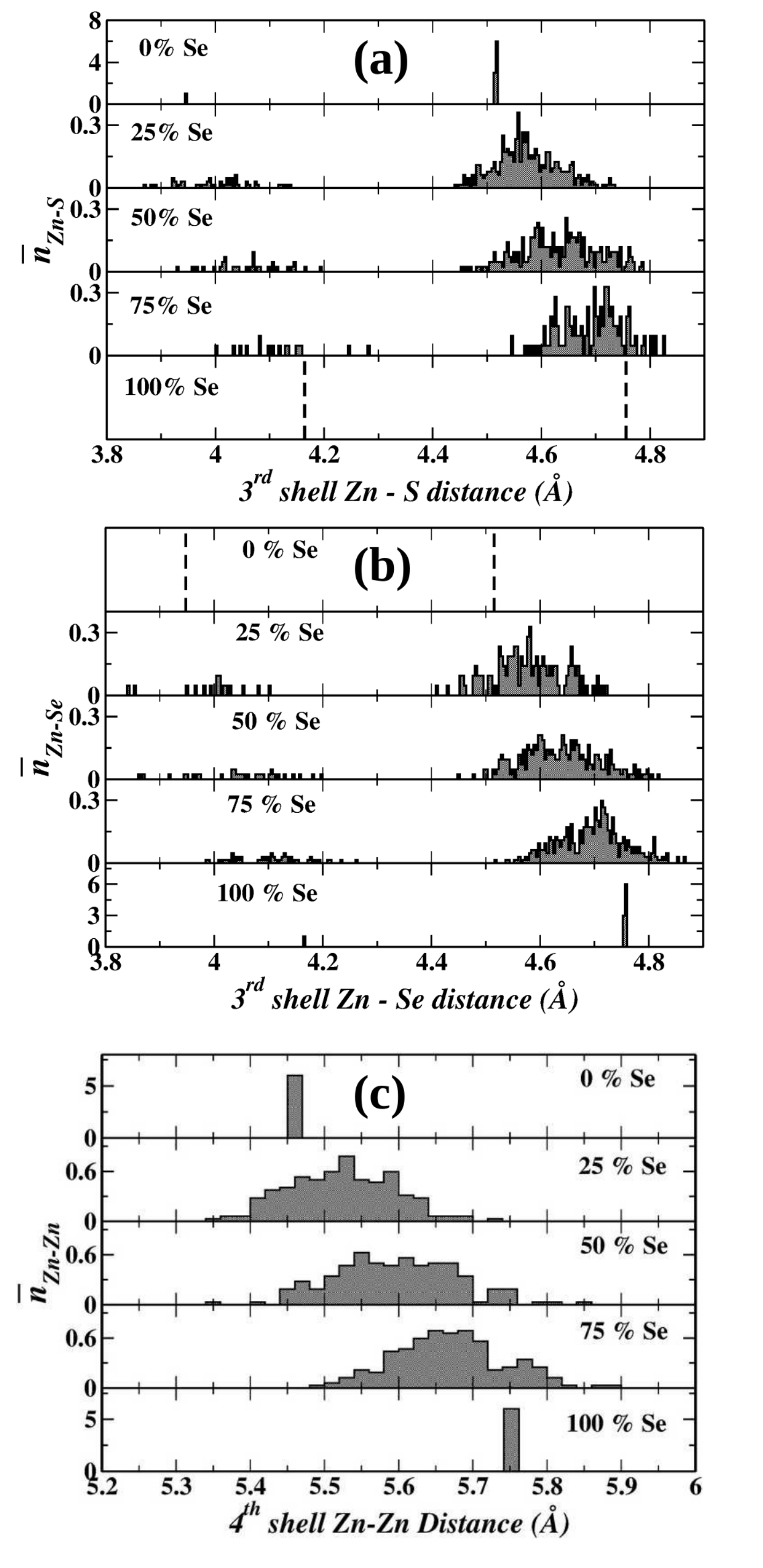}
  \caption{(a) Distributions of the third neighbor short and long Zn-S bonds in the solid solution alloy for increasing Se concentration (top to bottom), normalized with respect to the number of bonds per Zn atom (10). The two vertical dashed lines in the bottom panel represent the corresponding values in pristine ZnSe.  (b) Distributions of the third neighbor short and long Zn-Se bonds in the solid solution alloy for increasing Se concentration (top to bottom), normalized with respect to the number of bonds per Zn atom (10). The two vertical dashed lines in the top panel represent the corresponding values in pristine ZnS. (c) Distributions of the fourth neighbor Zn-Zn bond distances in the system with increasing Se concentration (top to bottom), normalized with respect to the number of bonds per Zn atom (6). More quantitative details on the range and averages of the distributions are provided in the SM~\cite{SM}.}
  \label{Fig7} 
\end{figure}

The variation of the average second neighbor $X-X$ ($X$ = S or Se) bond lengths reported in Fig.~\ref{Fig3}(b) shows a behavior resembling what one would expect from Vegard's law. Nevertheless, some deviations from it are still visible and therefore one may ask to what extent this behavior persists into the next shells of neighbors. 
In the case of pure ZnS/ZnSe in the wurtzite structure, the third shell around each Zn atom consists of 10 S/Se atoms, divided in three different types by symmetry (see Table~\ref{TableA1}). Two of those types have roughly the same bond lengths, which means that we expect to see only two different bond lengths. For example, the Zn-S bond in pristine ZnS has 1 short value of about 3.95 \r{A} and 9 larger values of about 4.51 \r{A}.  Variations in the average value of short and long Zn-S (circle and square) and Zn-Se (down and up triangles ) bonds between Zn and S/Se atoms in the third coordination shell are illustrated in Fig.~\ref{Fig6}(a).
The black dashed line indicates the variation of a single Zn-V bond between Zn and a composition averaged virtual atom (V) replacing both S and Se atoms, which would be predicted by VCA. Looking at short and long bonds separately, the three curves are on top of each other, which clearly shows that the third neighbor bond lengths follow Vegard's law more accurately compared to those of the first and second neighbors. 
We stress again that the origin of these different bond lengths is not a different local environment, as in Fig.~\ref{Fig2}(a) or Fig.~\ref{Fig3}(a), but simply the result of the crystal symmetry of the wurtzite phase. This feature is, in fact, absent in the data for the cubic phase, where 12 equidistant anions are present in the third coordination shell around each Zn atom (see Table~\ref{TableA1}). The histograms showing the distributions of the Zn-S and Zn-Se bond-lengths are reported in Figs.~\ref{Fig7}(a) and~\ref{Fig7}(b). The two separate regions corresponding to the two different bond lengths in the third coordination shell are evident. For the end members, we can also observe the split of the long bonds into two types, with 3 and 6 elements. Overall, no significant variation is observed between the bonds Zn-$X$, depending on choice of $X$. The distributions have now a substantial width, spreading over about 0.3 \r{A}. A similar spread is also observed in the zinc blend structure, as reported in the SM~\cite{SM}. It is somehow expected that the differences between wurtzite and cubic structures become less clear for pairs of atoms that are farther apart.

\FloatBarrier

At last, we also investigate the variation in the atomic bond lengths arising from the fourth shell of neighbors. This shell is particularly interesting for our analysis, since no experiment has been able to probe these longer distances. The analysis of the trends reported for the first three shells suggests that a behavior in accordance with Vegard's law should be expected. The data for the average Zn-Zn bond lengths as a function of increasing Se concentration, shown in Fig.~\ref{Fig6}(b), confirm this hypothesis. This establishes firmly that any technique probing over distances above the third shell of neighbors will provide results showing a linear trend with respect to the composition, at least for $\text{ZnSe}_{x}\text{S}_{1-x}$. This is the situation typical of diffraction experiments, which are used to affirm Vegard's law. As expected, a single mode characterizes the distributions of the fourth neighbor bond lengths, shown in the histograms of Fig.~\ref{Fig7}(c). However, a substantial width of the distribution is still visible, which raises the interesting question on what coordination shell will show a narrow distribution that is indistinguishable from the behavior of the lattice constant. We compared the histograms of the distributions of farther and farther neighbors and observed a slow decrease of the width (data not shown). However, our analysis is probably limited by the size of our supercells, which provide an exact match of the pair correlation function of a random alloy only up to the $7^\text{th}$ coordination shell. Investigating larger supercells should provide a more detailed answer to this question, but this analysis is out of the scope of the present manuscript.

 \subsection{A model of the changes at the atomic scale due to doping} 

If we consider the behavior of the first nearest neighbor Zn-S and Zn-Se bond lengths, as well as the behavior of the second neighbor bond lengths and their bond angles, we can construct a model for the structural changes at the atomic scale taking place in the anion substituted $\text{ZnSe}_{x}\text{S}_{1-x}$ alloy system in either wurtzite/cubic phase~\cite{SMukherjee_thesis_2018,TDan_thesis_2020}.  In these alloy systems, where the substitution is done at the anionic sublattice, the SZn$_4$ and SeZn$_4$ tetrahedral units act as the basic building blocks and are connected via a corner shared Zn atom.
This is a consequence of the almost invariant nature of the local Zn-S and Zn-Se bonds and anion-centered bond angles. Now, as a result of their different sizes, the SZn$_4$ and SeZn$_4$ units rotate/tilt with respect to each other, to accommodate and reduce the local strain. This results in a change of the $X$-Zn-$X$ bond angles, which we know, from Fig.~\ref{Fig5}(a), to vary as a function of composition. Thus, this leads to a much rapid increase or decrease in the $X-X$ second neighbor bond lengths, whose behavior becomes closer to what a microscopic analogue of Vegard's law would predict.

 \begin{figure}[ht]
\includegraphics[scale=0.8]{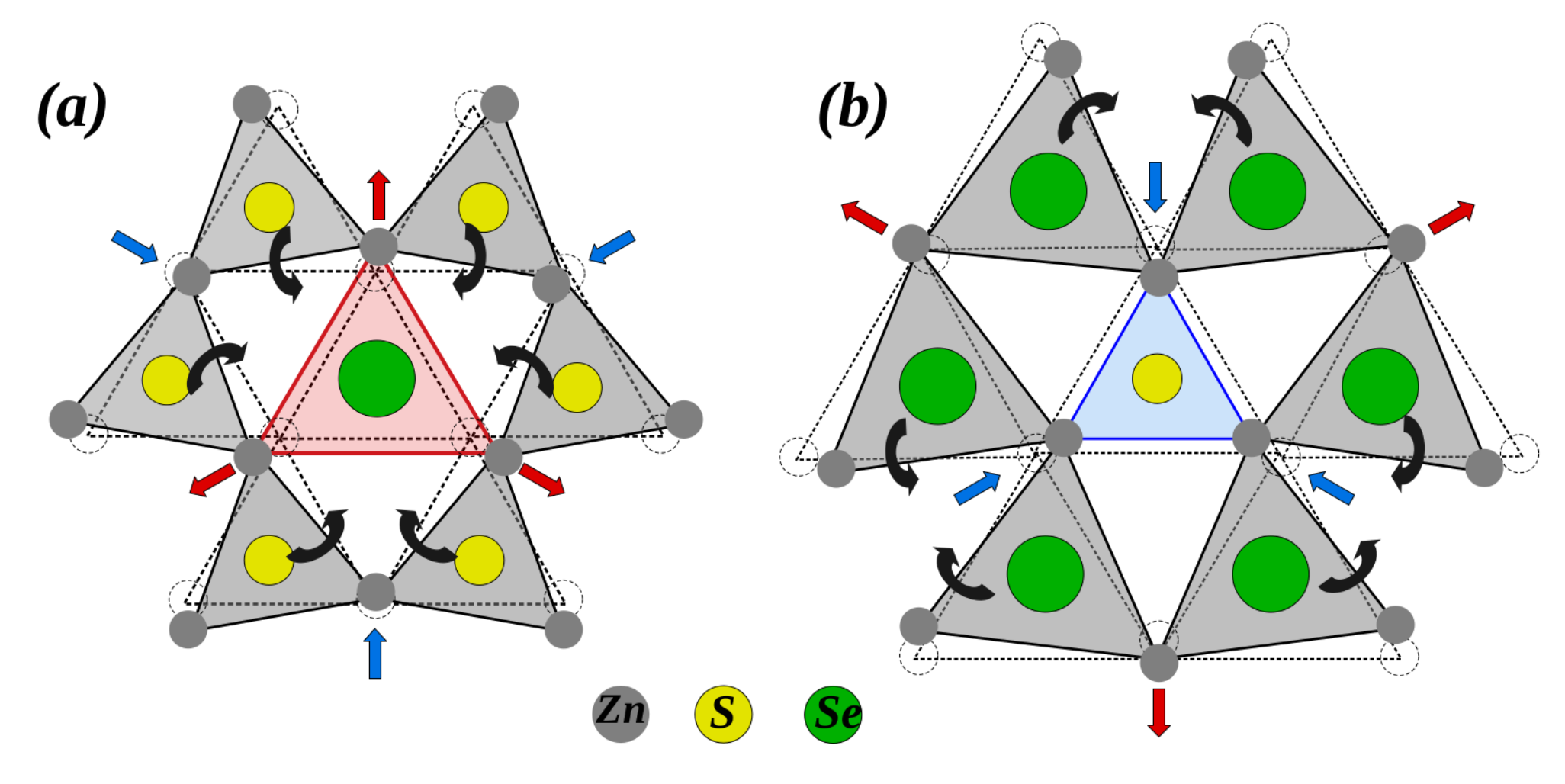}
\caption{(a) Substitution of a Se atom in a S rich region leading to a larger SeZn$_4$ tetrahedral unit (red triangle) and rotation of the surrounding SZn$_4$ tetrahedral units shown by black curved arrows. (b) Substitution of a S atom in a Se rich region leading to a smaller SZn$_4$ tetrahedral unit (blue triangle) and rotation of the surrounding SeZn$_4$ tetrahedral units shown by black curved arrows. }
\label{Fig8}
\end{figure}

The previous analysis allows us to model the details of the two extreme situations, where we substitute a single Se atom in an S-rich region and vice versa. 
Substituting a Se atom in an S-rich region is schematized in Fig.~\ref{Fig8}(a). The larger volume of the Se atom (green sphere) leads to a larger SeZn$_4$ tetrahedral unit, as indicated by the red triangle. The nearest neighbor Zn atoms (gray spheres) around the Se atom are displaced from their ideal positions, causing the associated S-Zn-S angles to increase. This has been shown by the small red arrows. Such movements of the Zn atoms further lead to a rotation of the surrounding SZn$_4$ tetrahedral units (gray triangles) as a result of the chemical pressure, depicted through black curved arrows. These cooperative rotations allow the S atoms (yellow spheres) inside the tetrahedral units to come closer to the Se atom, causing the S-Zn-Se bond angles to decrease.  The initial state of the tetrahedral units before the substitution of the Se atom at the S site is indicated by the black dotted lines. 
An analogous scheme can be drawn for the situation where an S atom is substituted in a Se-rich region, shown in Fig.~\ref{Fig8}(b). 
A smaller SZn$_4$ tetrahedral unit (blue triangle) leads to a decreased Se-Zn-Se bond angles (indicated by the small blue arrows), forcing the surrounding SeZn$_4$ tetrahedral units to rotate in such a manner that the Se-Zn-S bond angles increase. 
So, from these two extreme pictures we understand that for an increasing Se concentration the S-Zn-S/S-Zn-Se bond angles shall gradually increase/decrease - see Fig.~\ref{Fig5}(a) - which leads to a more rapid increase in the second neighbor S-S distance with respect to the first neighbors, in agreement with Fig.~\ref{Fig3}(b). Similarly, for an increasing S concentration, the Se-Zn-Se/Se-Zn-S bond angles shall gradually decrease/increase - see Fig.~\ref{Fig5}(a) - ensuring a rapid decrease in the Se-Se second neighbor distance with increasing S concentration, in agreement with Fig.~\ref{Fig3}(b).
From these structural considerations at the atomic scale, we can also observe an interesting competition in the structural relaxation. An increase in the S-Zn-S bond angle favors a decrease of the S-Zn-Se bond angle, whereas a decrease in the  Se-Zn-Se bond angle favors an increase of the S-Zn-Se bond angle (we consider the S-Zn-Se and Se-Zn-S angles to be coincident here). These competing mechanisms lead to average values of the S-Zn-Se angle that lie in the region between the S-Zn-S and Se-Zn-Se values, exactly as depicted in Fig.~\ref{Fig5}(a).
Furthermore, as the rotations of the tetrahedral units take place around the S/Se atoms, we expect the Zn-$X$-Zn (X = S or Se) bond angles to remain invariant when the the composition is changed, as visible in Fig.~\ref{Fig5}(b).
This reveals unambiguously that the SZn$_4$ and SeZn$_4$ tetrahedral units try to constitute the basic and probably rigid building blocks of the alloy systems.
Similar considerations based on the bond lengths and bond angles can be made for the cation substituted $\text{Zn}_x\text{Cd}_{1-x}\text{S}$ alloy system~\cite{SMukherjee_PRB_2014} in the wurtzite phase, suggesting that the ZnS$_4$ and CdS$_4$ tetrahedral units can be identified as the basic building blocks. 
Hence, these observations from two very different alloy systems suggest that, in general, one cannot consider either the cation or anion-centered tetrahedral units as the basic building blocks. 
In the pristine systems, both units are equivalent. The difference between them occurs only through the alloying process that promotes one over the other as a basic unit, based on the nature of the substitution (cationic/anionic).

 \begin{figure}[ht]
\includegraphics[scale=0.5]{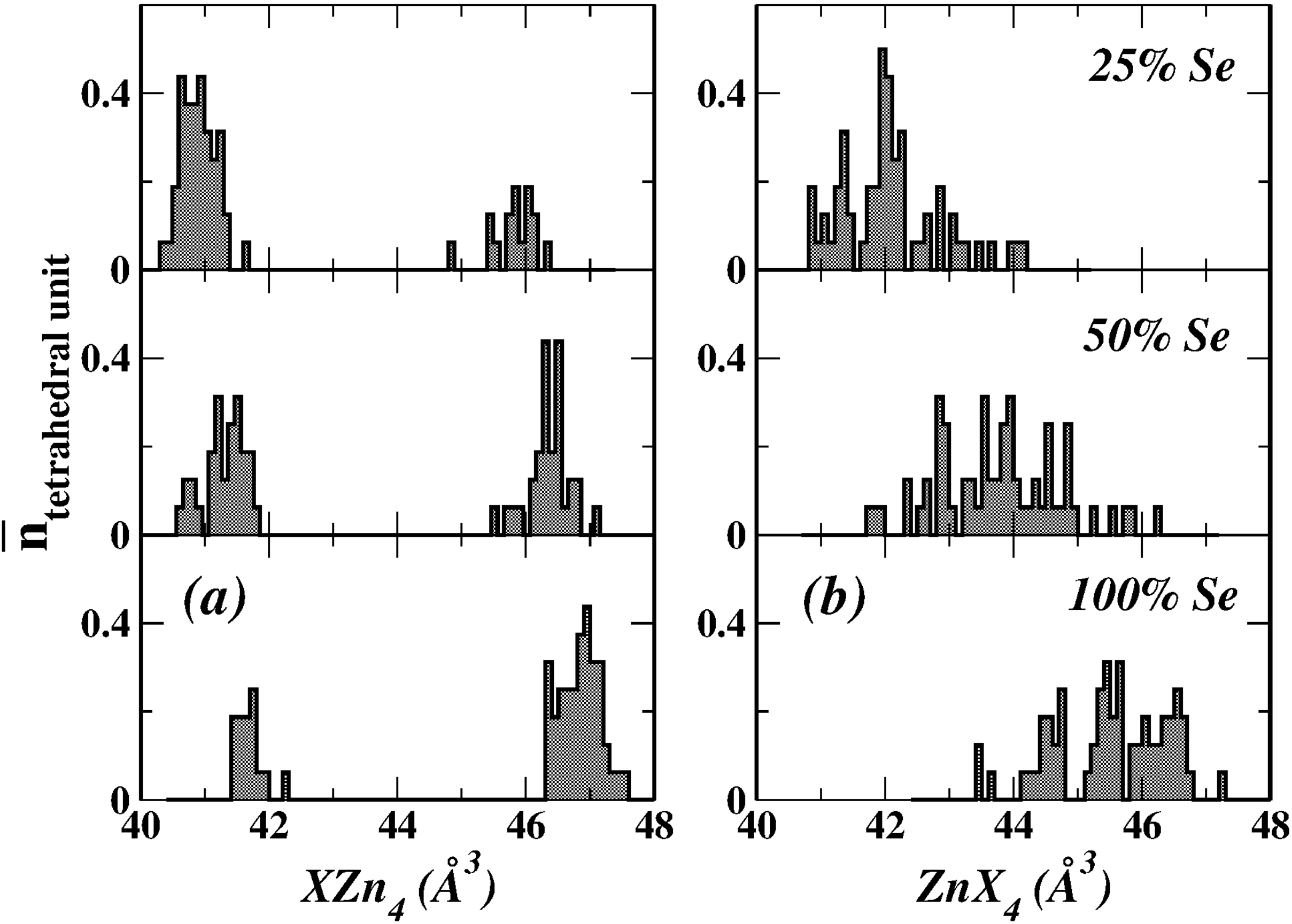}
\caption{Distribution of the calculated volumes associated with the (a) $X$Zn$_4$, and (b) Zn$X_4$ units (X = S or Se) in the alloy systems, normalized to 1.}
\label{Fig9}
\end{figure}

After having identified cation or anion-centered tetrahedral units as the basic building blocks in the alloy systems, we can proceed to analyze their effective rigidity. Here, by rigidity we mean that, in the case of the anion substituted system, the SZn$_4$ and SeZn$_4$ units can maintain their tetrahedral shapes in the alloy systems/solid solutions. 
This rigidity has been suggested in previous literature~\cite{SMukherjee_PRB_2014,DD2}, on the basis of  the variation of the average values of bond lengths and bond angles. 
However, the aforementioned analysis ignores the total picture that we get from their corresponding distributions and therefore may be misleading. 
As we described above, the variation of the average values can be understood from the picture of substituting a S/Se atom in the ZnSe/ZnS host in the dilute limit, but things can be very different for a realistic doping concentration, where disorder plays an important role. 
Fig.~\ref{Fig3}(c) illustrates clearly that, for a 50\% concentration of Se, the Zn-Zn second neighbor bonds have a very broad distribution that cannot really be interpreted as bimodal. A similar absence of bimodality can be inferred from the analysis of the Zn-S-Zn or Zn-Se-Zn bond angles, as visible from Fig.~\ref{FigB2} in Appendix B. 
This observation is important, as the short and long Zn-Zn bonds construct the tetrahedral units around the S and Se atoms respectively. 
Hence, the absence of bimodality in these distributions suggest that the SZn$_4$ and SeZn$_4$ units are not able to preserve their tetrahedral shapes in the alloy systems and cannot be considered as rigid.
So, if they are not rigid enough to preserve their tetrahedral shapes, the next question is whether the SZn$_4$ and SeZn$_4$ units preserve their volumes or not. 
To answer this question, we calculated the volumes associated with all the  SZn$_4$ and SeZn$_4$ units present in the system. Their distribution is illustrated in Fig.~\ref{Fig9}(a) for the three alloy systems we considered here. 
Since a clear bimodal distribution is observed, the SZn$_4$ and SeZn$_4$ units can be described as not so rigid, but volume preserving. 
The presence of volume-preserving distortions also explains the invariance of the average anion-centered bond angles, reported in Fig.~\ref{Fig5}(b).
A very different scenario is observed for the cation-centered ZnX$_4$ units, which not only are non-rigid but do not even preserve their volumes.
This is clearly visible in the histogram of the calculated volumes of the ZnX$_4$ units, shown in Fig.~\ref{Fig9}(b). 
The broad distribution observed in this case is most likely due to the fact that ZnX$_4$ units have five possible variants, ZnS$_4$, ZnS$_3$Se$_1$, ZnS$_2$Se$_2$, ZnS$_1$Se$_3$, and ZnSe$_4$, with all of them having different volumes. 
In conclusion, in this kind of alloy systems, the nearest neighbor cation-anion bonds always try to retain their value close to the values observed in the pure end members. The local chemical pressure resulting from the disordered substitution of a different atom at a given lattice site is compensated by a change in the bond angles leading to a rotation and tilt of the tetrahedral units that constitutes the basic building block of the system, which are SZn$_4$ and SeZn$_4$ units in our case. 
In the case of alloying at the cationic sublattice, cation-centered units act as the basic building blocks.

\subsection{Electronic structure and band gap}

While Vegard’s law predicts a linear behavior for the evolution of the lattice parameters in a solid solution alloy, experimental measurements show a quadratic behavior for the band gap.  The deviation from linearity is associated to a bowing parameter $b$, whose size increases when the difference in band gap between the end members increases~\cite{JEBernard_PRB_1987,JWu_SSC_2003}. During the years, there have been a few attempts to provide a formula connecting the bowing parameter to relevant microscopic quantities, such as the mismatch in size, band gap or electronegativity, but only with limited success~\cite{JEBernard_PRB_1987,JWu_SSC_2003,Mader_PRB_1995,Wei_PRB_1997,Shan_PRL_1999,Kent_PRB_2001,Ferhat_PRB_2002,Wu_PRB_2003}. In this section, we will perform a systematic analysis of various effects contributing to the bowing parameter, in the attempt of identifying what mechanisms play the most crucial role in $\text{ZnSe}_{x}\text{S}_{1-x}$. 

We start by focusing on the atom-projected density of states for pristine ZnS and ZnSe in their DFT wurtzite optimized structure, reported in Fig.~\ref{FigC1}. The valence band is mainly formed by S/Se-$p$ states, plus smaller contributions due to Zn-$s$ and Zn-$p$ states. 
The conduction band is, instead, constituted in equal measure of Zn-$s$, Zn-$p$ and S/Se-$p$ states. 
Overall, Fig.~\ref{FigC1} is in excellent agreement with previous reports~\cite{ATorabi_JPCL_2015,Hussain2019MT,SSapra_PRB_2002,RViswanatha_PRB_2005,SDatta_JPCM_2008}. 
Since the band gap is fully specified by Zn-$s$ and S/Se-$p$ states, we can focus on those for the analysis of the evolution of the density of states in the $\text{ZnSe}_{x}\text{S}_{1-x}$ alloy, illustrated in Fig.~\ref{FigC2}. 
We can see that doping does not induce any new states, but the band edges move continuously across the various compositions. 
This behavior is consistent with the typical experimental observations on isovalent alloys, although there are a few exceptions to this rule, as e.g. $\text{GaP}_{x}\text{N}_{1-x}$ and $\text{GaP}_{x}\text{Bi}_{1-x}$~\cite{JEBernard_PRB_1987}. 
To further verify that this picture holds in $\text{ZnSe}_{x}\text{S}_{1-x}$, we performed additional calculations for a single Se (S) impurity in a 128 atom supercell of ZnS (ZnSe). These results (data not shown) confirm that no detached impurity level forms, even in the dilute limit.

\begin{figure}[h]
\includegraphics[scale=0.5]{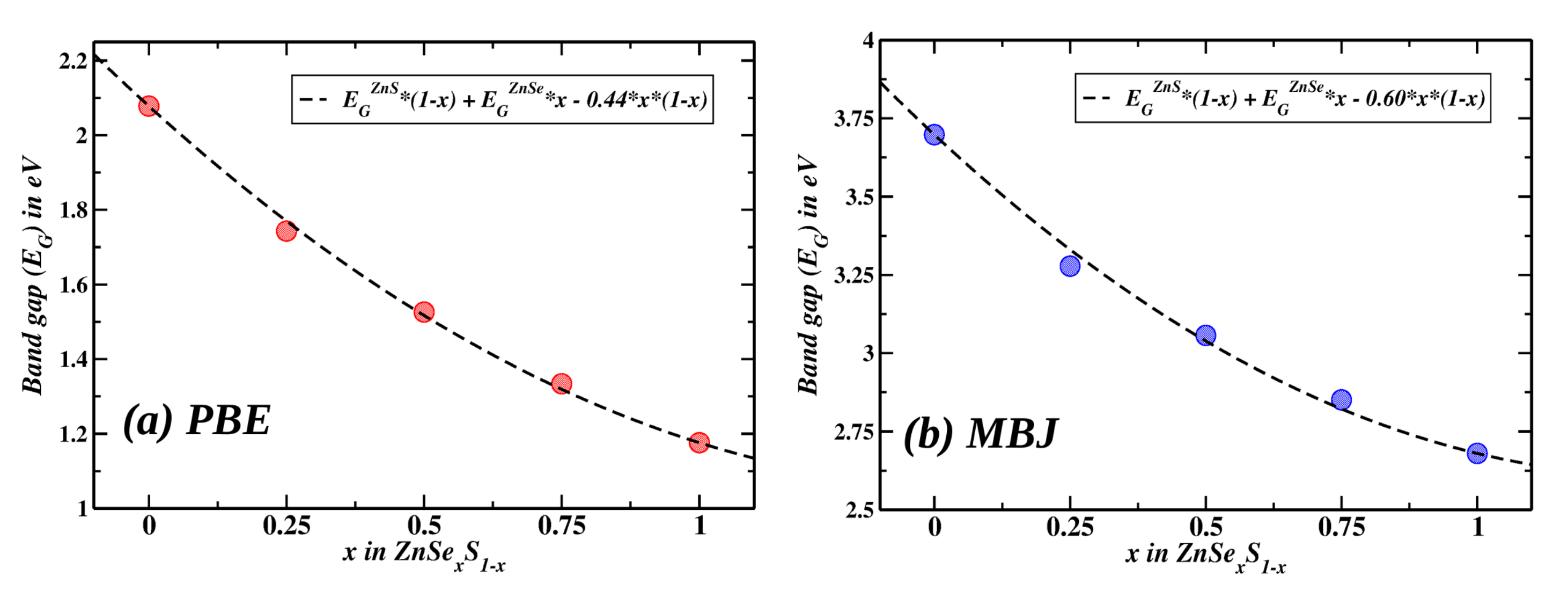}
\caption{Variation in the band gap as a function of increasing Se concentration calculated using (a) PBE  and (b) mBJ as exchange correlation functional. The black dotted line represents a quadratic fit, which is used to extract the bowing parameter $b$. As reported in the legends, $b=0.44$ eV for PBE and $b=0.60$ eV for mBJ.}
\label{Fig10}
\end{figure}

The aforementioned analysis of the spectra can be used to extract the band gap E$_g$. The values for pristine ZnS and ZnSe in their DFT optimized wurtzite structure amount to 2.08 and 1.18 eV, respectively. The experimentally reported band gap of ZnS in the wurtzite structure is in the range of 3.86 - 3.98 eV, whereas for ZnSe measurements indicate the value of about 2.87 eV~\cite{MOshikiri_PRB_1999,Semi_Springer_1992}. 
This strong discrepancy between theory and experiment is due to the usage of a semi-local exchange-correlation functional, as the GGA-PBE, as already emphasized in previous works~\cite{ATorabi_JPCL_2015,Hussain2019MT,SSapra_PRB_2002,RViswanatha_PRB_2005,SDatta_JPCM_2008}. 
As shown below, a more accurate evaluation of the band gap can be obtained via more complex functionals, as e.g. meta-GGA or hybrid functionals~\cite{ATorabi_JPCL_2015}.
The relative variation of the band gap as a function of increasing Se content in the $\text{ZnSe}_{x}\text{S}_{1-x}$ solid solution is shown in Fig.~\ref{Fig10}(a). The deviation from linearity is evident, and fitting the data points to a quadratic equation leads to a bowing parameter \textit{b} = 0.44 eV. 
This value is in very good agreement with previously reported theoretical values, for example, with the value of 0.46 eV reported in Ref.~\cite{DLong_CMS_2018} by using a $\Delta$-Sol correction to DFT in GGA. 
A good agreement is also observed with respect to reported experimental values, which are in the range of 0.35 - 0.68 eV~\cite{JEBernard_PRB_1987,JLu_Nanoscale_2012,AEbina_PRB_1974,MAAviles_InoChem_2019}. This large interval reflects variations in sample homogeneity and the phase admixture between wurtzite and zinc blende that may depend on the S/Se concentration~\cite{MAAviles_InoChem_2019}. 
To verify if the error on the precise size of the band gap affects the bowing parameter significantly, we performed additional calculations with the mBJ exchange-potential in combination with LDA correlations~\cite{MBJ_1,MBJ_2}.
The predicted band gaps for pristine ZnS and ZnSe in the GGA-PBE optimized wurtzite structure amount to 3.70 and 2.68 eV, respectively. As expected, the agreement with the experimental values is significantly improved with respect to PBE. The variation of the band gap for increasing Se concentration is reported in Fig.~\ref{Fig10}(b). 
Fitting the data points to a quadratic equation leads to a bowing parameter \textit{b} = 0.60 eV. 
This demonstrates that the choice of the exchange-correlation functional does not only affect the values of the band gap across the full range of compositions, but also changes the bowing parameter itself.

\begin{figure}[h]
\includegraphics[scale=0.5]{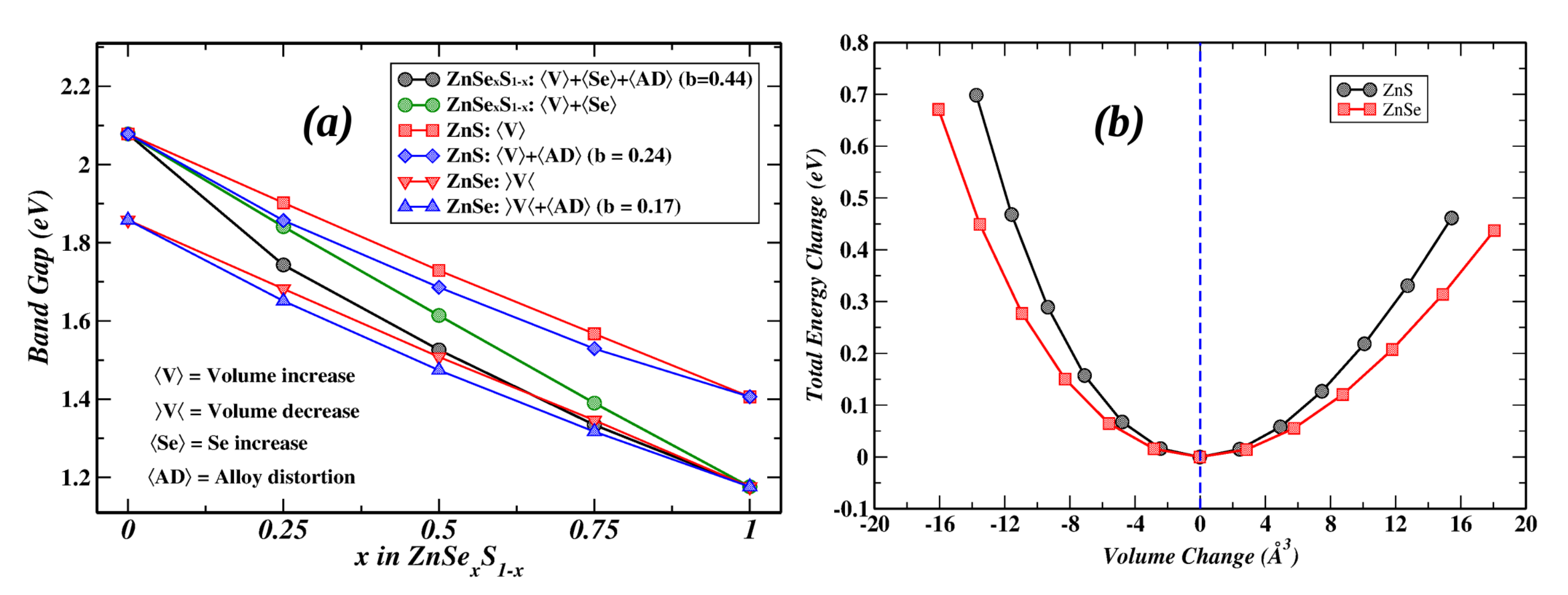}
\caption{(a) Decomposition of various mechanisms contributing to the bowing parameter. As indicated in the legend, volume changes and structural distortions as a result of substitution in the $\text{ZnSe}_{x}\text{S}_{1-x}$ alloy, pure ZnS, and ZnSe are considered. See main text for comprehensive details. (b) Total energy vs atomic volume curves for ZnS and ZnSe in the wurtzite phase. Zero volume corresponds to the equilibrium atomic volume of each compound, for a better comparison of the relative stiffness. }
\label{Fig11}
\end{figure}

A better understanding of the nature of the bowing parameter in $\text{ZnSe}_{x}\text{S}_{1-x}$  can be achieved by identifying the most important  contributions to it. For computational convenience, this analysis will be based on the GGA-PBE calculations, but our conclusions are most likely valid for other DFT functionals as well, as discussed at the end of this section.
Our current understanding of the evolution of the band gap across different concentrations in binary alloys is based on the theoretical work of Ref.~\onlinecite{JEBernard_PRB_1987}, where the authors separate the effects due to volume change, electronegativity and structural deformation. The two steric mechanisms are those that we have already analyzed in the previous section, see e.g. Fig.~\ref{Fig8}. Starting from a ZnS environment, if we gradually replace S with Se, the larger size of the latter will drive a volume change of the tetrahedral building blocks. Moreover, the disorder of the dopants will induce a volume preserving structural distortion of the tetrahedra (see Section III.D), changing the cation-anion-cation or anion-cation-anion bond angles. These geometrical changes will modify the local electronic structure, by reducing the band gap. The third effect is associated directly to the increased electronegativity of Se with respect to S. 
In Ref.~\onlinecite{JEBernard_PRB_1987}, all these mechanisms were analyzed by extracting their corresponding contribution from a few limited calculations, at a fixed concentration. However, due to the larger computational resources available today, we are able to calculate directly the evolution of the band gap for the various supercells under well-defined constraints. This analysis is illustrated in Fig.~\ref{Fig11}(a), where the black circles represent the fully self-consistent data already shown in Fig.~\ref{Fig10}(a). The green circles represent the results of calculations where the structural deformation beyond the change of volume due to Vegard's law is not allowed. The linear behavior shows unequivocally that the local structural deformation is the main factor contributing to the bowing parameter, while the remaining contributions, associated to volume deformation and  electronegativity, are very small.
This comparison is in agreement with the aforementioned studies, but provides also a new insight, i.e. the fact that the deviation from linearity arises mainly in the S-rich region, as shown by the more marked differences between black and green curves there.
This feature may appear obvious at first, since we expect an asymmetric behavior due to the fact that adding a building block whose local electronic structure has a smaller band gap with respect to its host (as in the ZnS limit) decreases the system band gap, while the opposite situation (as in the ZnSe limit) does not lead to any increase.
However, one may expect this asymmetry to be connected to the ionic nature of the substitution and hence it should not be fully canceled when structural deformations are suppressed. Since this asymmetry is not visible in the green curve, the main mechanism driving the asymmetry of the band gap bowing should have a different origin.

A better understanding of these data can be achieved by removing all the effects due to the different electronegativity. To this aim, we evaluate the band gaps of pure ZnS and ZnSe in the perfect wurtzite structure, but using lattice parameters obtained from Vegard's law across the various concentrations.
These data are illustrated by the two red curves in Fig.~\ref{Fig11}(a). As expected, a perfectly linear behavior is observed.
We can now evaluate the role of the structural distortions while ignoring the chemical composition by calculating pure ZnS and ZnSe system in the structure corresponding to the optimized supercells of the solid solutions at each concentration. These data correspond to the two blue curves in Fig.~\ref{Fig11}(a). A substantial bowing is visible and can be quantified as $b$ equal to 0.24 eV and 0.17 eV for ZnS and ZnSe, respectively. We can see, then, that the same structural distortions in ZnS give rise to a larger bowing in comparison to ZnSe, albeit no asymmetry is noted with respect to the 50\% concentration. Hence, the fact that the bowing parameter arises mainly in the ZnS region can be explained as a coexistence of the previous two effects, namely the asymmetric effect of inducing a local electronic structure modification in a host with a given band gap and the larger response of the electronic structure of ZnS to equivalent structural modifications. This also indicates that structural distortion as well as the difference in the atomic properties of S and Se atoms are necessary to account for the actual bowing. Finally, the previous analysis raises the question why the same structural distortion gives rise to a slightly larger bowing in the ZnS system compared to ZnSe. To provide an answer, we calculated the energy vs volume curves for both ZnS and ZnSe, shown in Fig.~\ref{Fig11}(b). 
The curve for ZnS is clearly stiffer than the one for ZnSe, indicating that inducing the same amount of local structural distortions costs more energy in the former than in the latter. The nearest neighbor Zn-S bonds and their corresponding bond angles are hence more rigid with respect to nearest neighbor Zn-Se bonds and corresponding angles. 
A more quantitative comparison of the stiffness of the curves in Fig.~\ref{Fig11}(b) can be obtained by calculating the bulk modulus $B$. Fitting the data by means of a Birch-Murnaghan equation of state~\cite{Murnaghan,Birch} leads to values of 68.76 GPa for ZnS and 56.10 GPa for ZnSe. For reference, the experimentally reported values for the zinc blende phase amount to 77.1 GPa and 62.0 GPa, for ZnS and ZnSe respectively~\cite{THomann_SSS_2006,OMadelung_springer_1987}. 
These data demonstrate that a similar amount of structural distortion is going to result in a larger change in the interaction between the cationic and anionic orbitals in ZnS, if compared to ZnSe. This difference is then reflected in the variations of the electronic structure induced by said distortion.
Finally, we may speculate on how the presented analysis will depend on the choice of a different exchange-correlation functional. The bowing parameter is affected by the exchange-correlation functional directly, by modifying the band gap, and indirectly, by modifying the ionic positions. While the former effect may only bring quantitative changes, the latter may potentially change the decomposition illustrated in Fig.~\ref{Fig11}(a). However, this change is likely to be small, since GGA is very accurate in describing the structural properties, as shown by our analysis of pure ZnS and ZnSe. Previous calculations in LDA, where the largest discrepancy is expected, have corroborated this expectation by reaching the same conclusion as ours, i.e. that the local structural deformation is the main factor contributing to the bowing parameter~\cite{JEBernard_PRB_1987}. Analogously, the asymmetry of the bowing parameter with respect to the $50\%$ concentration is not supposed to be affected by a different functional. In fact, we demonstrated that this asymmetry is a consequence of having different bulk moduli for ZnS and ZnSe, in spite of their precise value. This will be qualitatively predicted by most functionals.

\section{Conclusions}
In summary, we investigated the structural and electronic properties of the solid solution $\text{Zn}\text{Se}_x\text{S}_{1-x}$ by means of density functional theory. The analysis was mainly focused on the wurtzite structure, but selected data for the cubic zinc blende structure were also discussed, when relevant. 
The quantum mechanical calculations were done directly via large supercells, generated as special quasi-random structures. Despite having a large computational cost, this approach has allowed us to investigate the role of structural and chemical disorder on the same footing, without averaging out significant effects, as e.g. may happen in VCA or CPA, due to their single-site nature. 
The structural analysis we performed showed that minor differences arise between the wurtzite and zinc blende structures, if an appropriate partitioning of the coordination shells is performed. These differences are mainly related to the width of the distributions of the bond lengths and arise from the different crystal symmetries affecting the distributions of the nearest neighbors.
The bond lengths in the first two coordination shells are characterized by bimodal distributions and hence inconsistent with what could be anticipated from the application of Vegard's law at the atomic level. In the third coordination shell, the bimodal character of the bond length distributions is lost, establishing a threshold for the validity of a microscopic Vegard's law. 
These features are even more evident for the fourth coordination shell, whose bond length distributions show one single peak and are virtually not distinguishable from the lattice constant.
Considering that no experiment has been able to probe this far, our findings provide an additional motivation for experimentalists to verify the validity of Vegard's law at the atomic scale. 
In this regard, additional information on the local structures can also be obtained with alternative experimental techniques, as e.g. the perturbed angular correlation (PAC) spectroscopy~\cite{Wichert_1995} or the nuclear magnetic resonance (NMR) and nuclear quadrupole resoance (NQR) spectroscopies~\cite{Haas_2017}.
Furthermore, the analysis of the bond lengths and angles across different coordination shells allowed us to identify tetrahedral building blocks around the anions and these are also shown to be non-rigid but volume preserving.  
Analogous building blocks are formed around the cations in case of cationic substitution in the solid solution, as e.g. for $\text{Zn}_x\text{Cd}_{1-x}\text{S}$. 
Finally, we connected the structural changes driven by the substitution to the bowing parameter that describes the evolution of the band gap in the solid solution. 
The deviation from the linearity was shown to arise mainly from structural deformation, which is in agreement with previous theoretical studies. 
The bowing of band gap is found to be asymmetric with respect to an equal concentration of S and Se, exhibiting the largest decrease across the solid solution in the S-rich region.
This behavior is shown to be connected to the difference in the stiffness of the two end members and, to a lesser extent, to the fact that the band gap is not an average quantity, but always defined by the closest lying single-particle excitations.

\section{Acknowledgments}
The computational work was enabled by resources provided by the Swedish National Infrastructure for Computing (SNIC) at UPPMAX, partially funded by the Swedish Research Council through grant agreement no. 2018-05973.
D.~D.~S. thanks the Science and Engineering Research Board, Department of Science and Technology, and the Council of Scientific and Industrial Research, Government of India, and Jamsetji Tata Trust for support of this research.
O.~E. and I.~D.~M. acknowledge financial support from the European Research Council (ERC), Synergy Grant FASTCORR, Project No. 854843.
I.~D.~M. and S.~S. acknowledge financial support from the National Research Foundation (NRF) of Korea, grant No. 2020R1A2C101217411, funded by the Korean government through the Ministry of Science and Technology Information and Communication (MSIT). The work of I.~D.~M. is also supported by the appointment to the JRG program at the APCTP through the Science and Technology Promotion Fund and Lottery Fund of the Korean Government, as well as by the Korean Local Governments, Gyeongsangbuk-do Province and Pohang City.

\appendix

\section{Analysis of neighbors around a Zn atom in ZnS Wurtzite}

The number of atoms, their types and the corresponding coordination shell as a function of increasing distance from a Zn atom in wurtzite and cubic ZnS are given in Table~\ref{TableA1}. We see that the fourth coordination shell around the Zn atom contains 6 Zn atoms at a distance of 5.45 \r{A}. In wurtzite ZnSe the distance is 5.74 \r{A}. These 6 Zn atoms are targeted for the analysis of the fourth shell environment. 

\setcounter{table}{0}
\renewcommand{\thetable}{A\arabic{table}}

\FloatBarrier

\begin{table}[]
\caption{Partition of the first five shells of nearest neighbors of a Zn atom in the GGA optimized ZnS. Both wurtzite and zinc blende structures are shown.}
\label{TableA1}
\begin{tabular}{|c|c|cc|cc|}
\hline
\multirow{2}{*}{Shell} & \multirow{2}{*}{Neighbor} & \multicolumn{2}{c|}{Wurtzite} & \multicolumn{2}{c|}{Cubic} \\ \cline{3-6} 
 &  & \multicolumn{1}{c|}{Number of atoms} & Distance & \multicolumn{1}{c|}{Number of atoms} & Distance \\ \hline
\multirow{2}{*}{First} & \multirow{2}{*}{S} & \multicolumn{1}{c|}{1} & 2.36 &  \multicolumn{1}{c|}{\multirow{2}{*}{4}} & \multirow{2}{*}{2.36} \\ \cline{3-4}
 & & \multicolumn{1}{c|}{3} & 2.36 & \multicolumn{1}{c|}{} &  \\ \hline

\multirow{2}{*}{Second} & \multirow{2}{*}{Zn} & \multicolumn{1}{c|}{6} & 3.85 & \multicolumn{1}{c|}{\multirow{2}{*}{12}} & \multirow{2}{*}{3.85} \\ \cline{3-4}
 & & \multicolumn{1}{c|}{6} & 3.86 & \multicolumn{1}{c|}{} &  \\ \hline

\multirow{3}{*}{Third} & \multirow{3}{*}{S} & \multicolumn{1}{c|}{1} & 3.95 & \multicolumn{1}{c|}{\multirow{3}{*}{12}} & \multirow{3}{*}{4.52} \\ \cline{3-4}
 & & \multicolumn{1}{c|}{3} & 4.51 & \multicolumn{1}{c|}{} &  \\ \cline{3-4}
& & \multicolumn{1}{c|}{6} & 4.51 & \multicolumn{1}{c|}{} &  \\ \hline

Fourth & Zn & \multicolumn{1}{c|}{6} & 5.45 & \multicolumn{1}{c|}{6} & 5.45 \\ \hline

\multirow{3}{*}{Fifth} & S & \multicolumn{1}{c|}{6} & 5.51 & \multicolumn{1}{c|}{\multirow{3}{*}{12}} & \multirow{3}{*}{5.94} \\ \cline{2-4}
 & S & \multicolumn{1}{c|}{6} & 5.93 & \multicolumn{1}{c|}{} &  \\ \cline{2-4}
 & S & \multicolumn{1}{c|}{3} & 5.95 & \multicolumn{1}{c|}{} &  \\ \hline

\end{tabular}
\end{table}
\FloatBarrier

\section{Histograms for bond angles distribution}

 \setcounter{figure}{0}
 \renewcommand{\thefigure}{B\arabic{figure}}

In this appendix we show the distribution of the anion and cation centric bond angles

\FloatBarrier

\begin{figure}[]

    \includegraphics[scale=0.5]{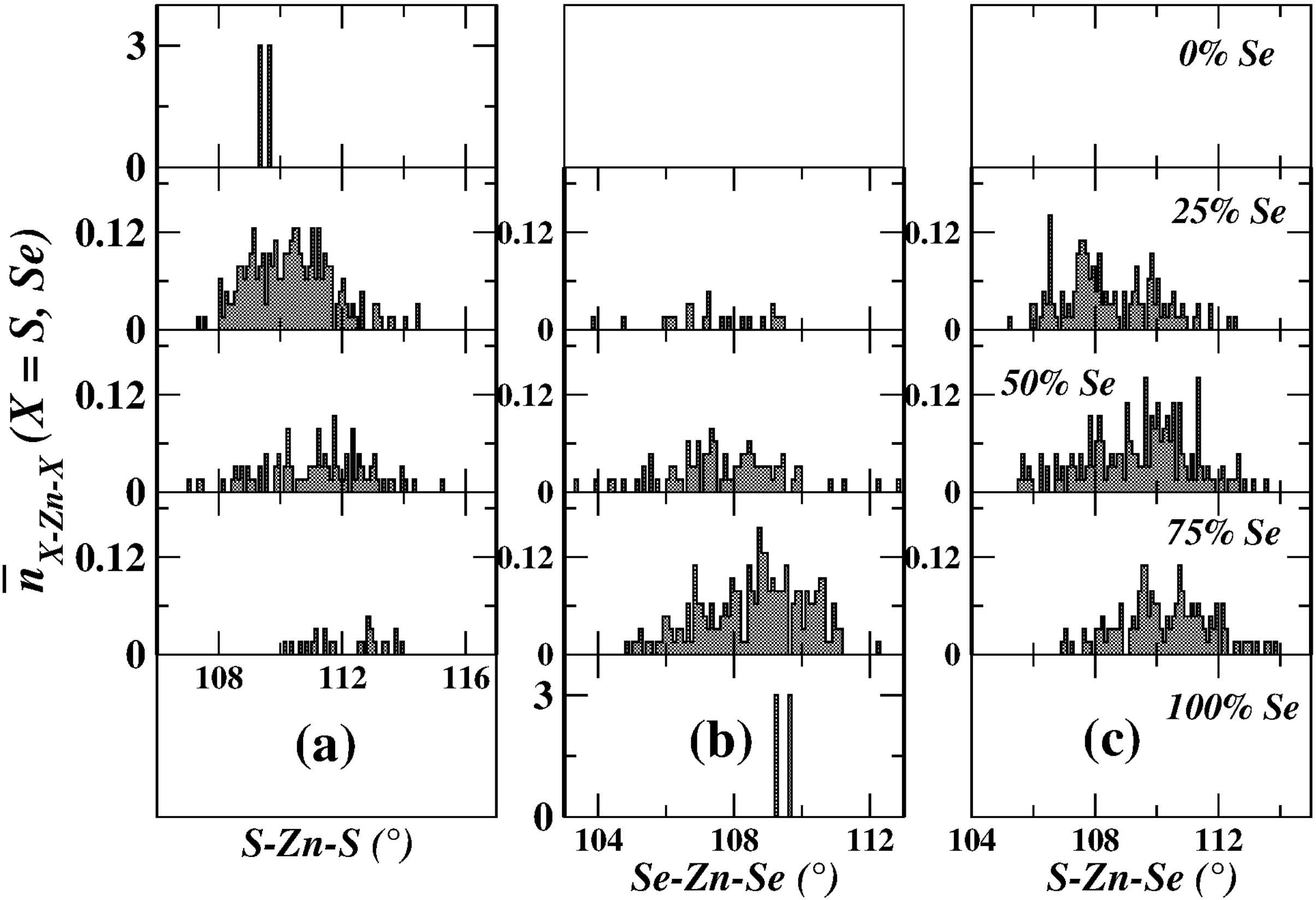}
  \caption{(a) Distributions of the S-Zn-S bond angles in the solid solution alloy for increasing Se concentration (top to bottom), normalized with respect to the number of bond angles centered on a Zn atom (6) weighted by the S concentration. (b) Distributions of the Se-Zn-Se bond angles in the solid solution alloy for increasing Se concentration (top to bottom), normalized with respect to the number of bond angles centered on a Zn atom (6) weighted by the Se concentration. (c) Distributions of the S-Zn-Se bond angles in the solid solution alloy for increasing Se concentration (top to bottom), normalized with respect to the number of bond angles centered on a Zn atom (6) weighted by the pair concentration of S and Se. The details of the average and range for these angles are given in given in the SM~\cite{SM}.}
  \label{FigB1}
\end{figure}

\begin{figure}[]

    \includegraphics[scale=0.5]{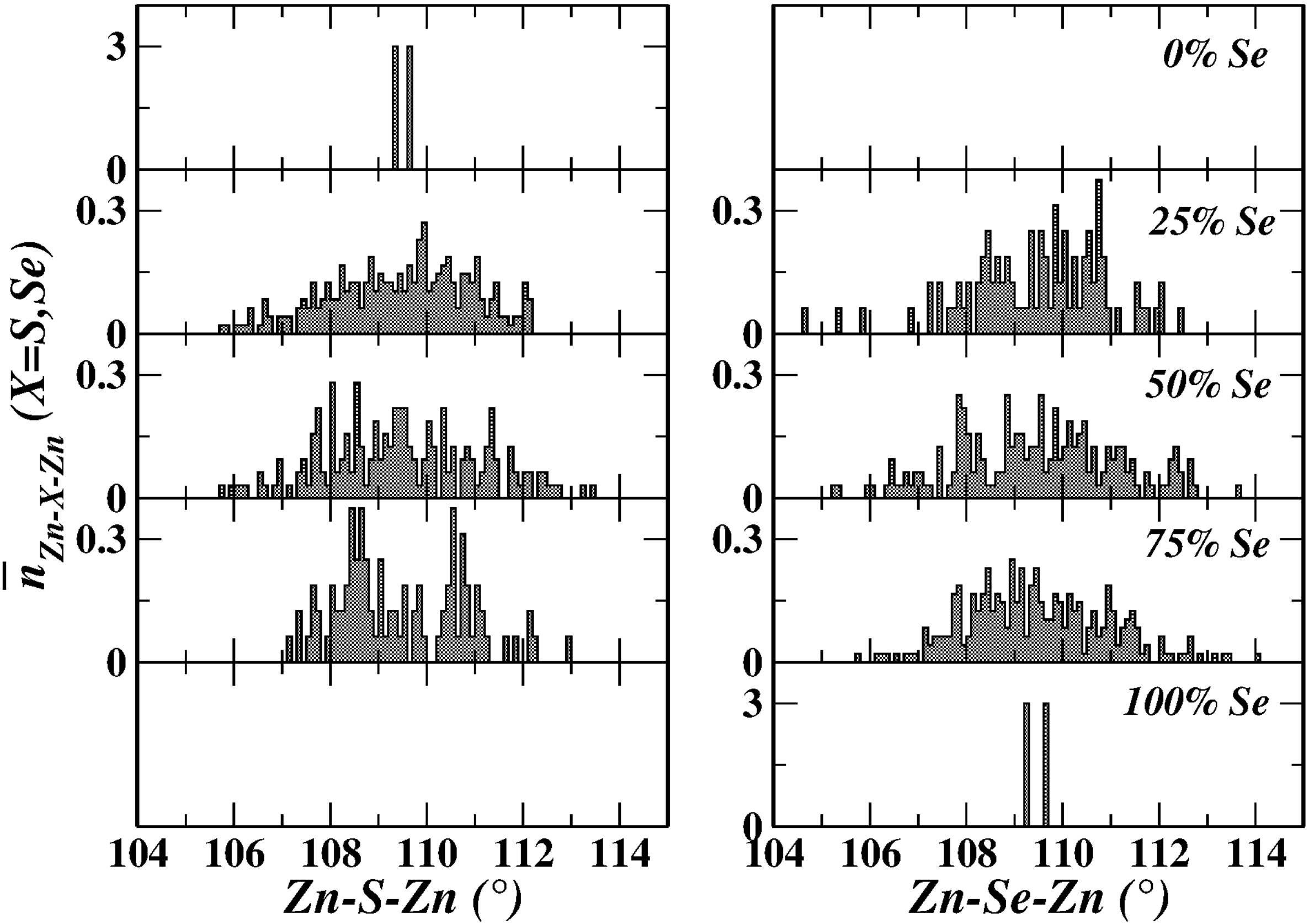}
  \caption{(a) Distributions of the Zn-S-Zn bond angles in the solid solution alloy for increasing Se concentration (top to bottom), normalized with respect to the number of bond angles centered on a S atom (6). (b) Distributions of the Zn-Se-Zn bond angles in the solid solution alloy for increasing Se concentration (top to bottom), normalized with respect to the number of bond angles centered on a Se atom (6). The details of the average and range for these angles are given in given in the SM~\cite{SM}.}
  \label{FigB2}
\end{figure}

\FloatBarrier

\section{Spectral properties}

 \setcounter{figure}{0}
 \renewcommand{\thefigure}{C\arabic{figure}}

\FloatBarrier
 \begin{figure}[ht]
\includegraphics[scale=0.5]{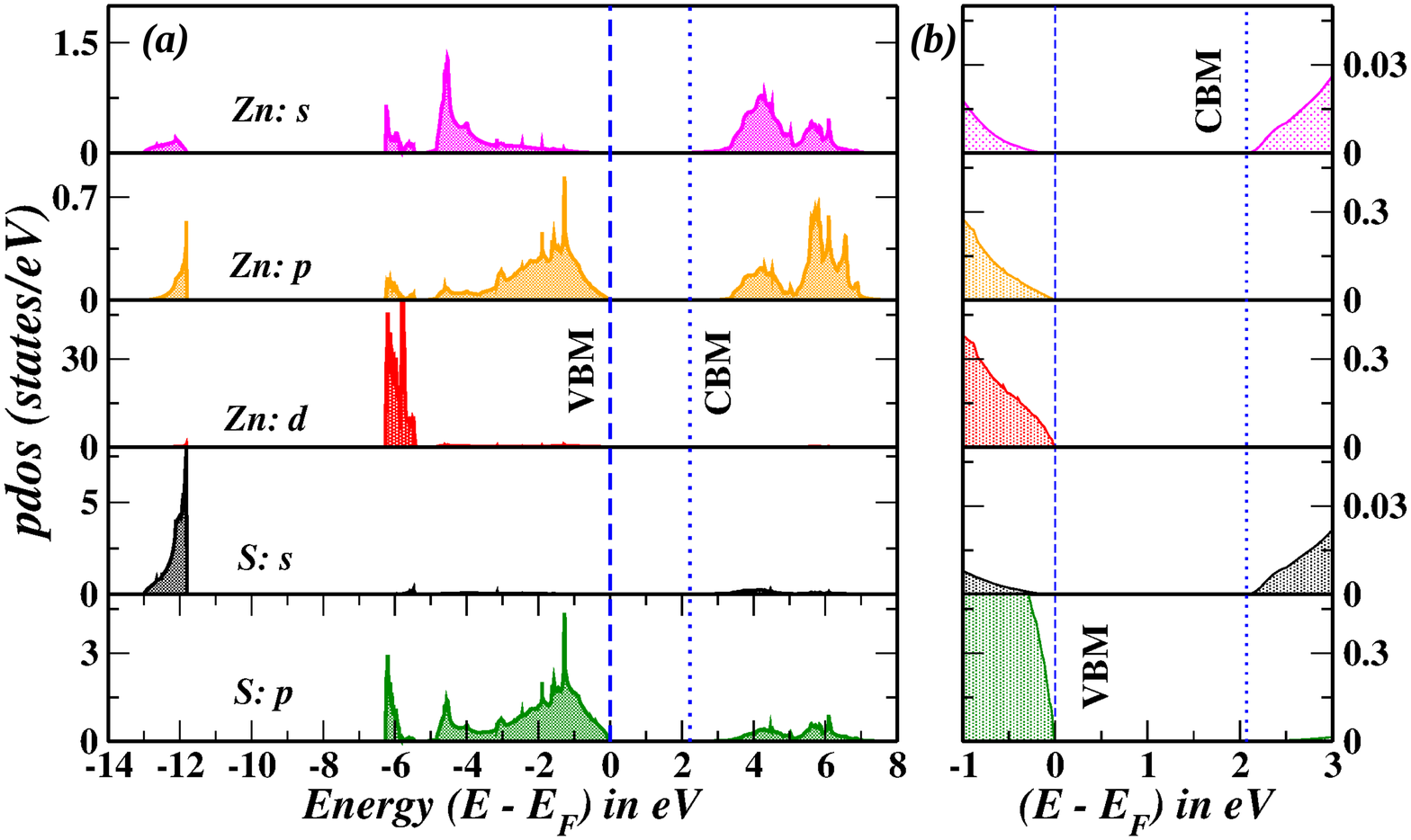}
\caption{(a) Atom projected partial density of states corresponding to the Zn and S states in ZnS. (b) Magnified region around the Fermi energy to show the contributions at the valence band maximum (VBM) and conduction band minimum (CBM). }
\label{FigC1}
\end{figure}
\FloatBarrier

\FloatBarrier
 \begin{figure}[ht]
\includegraphics[scale=0.5]{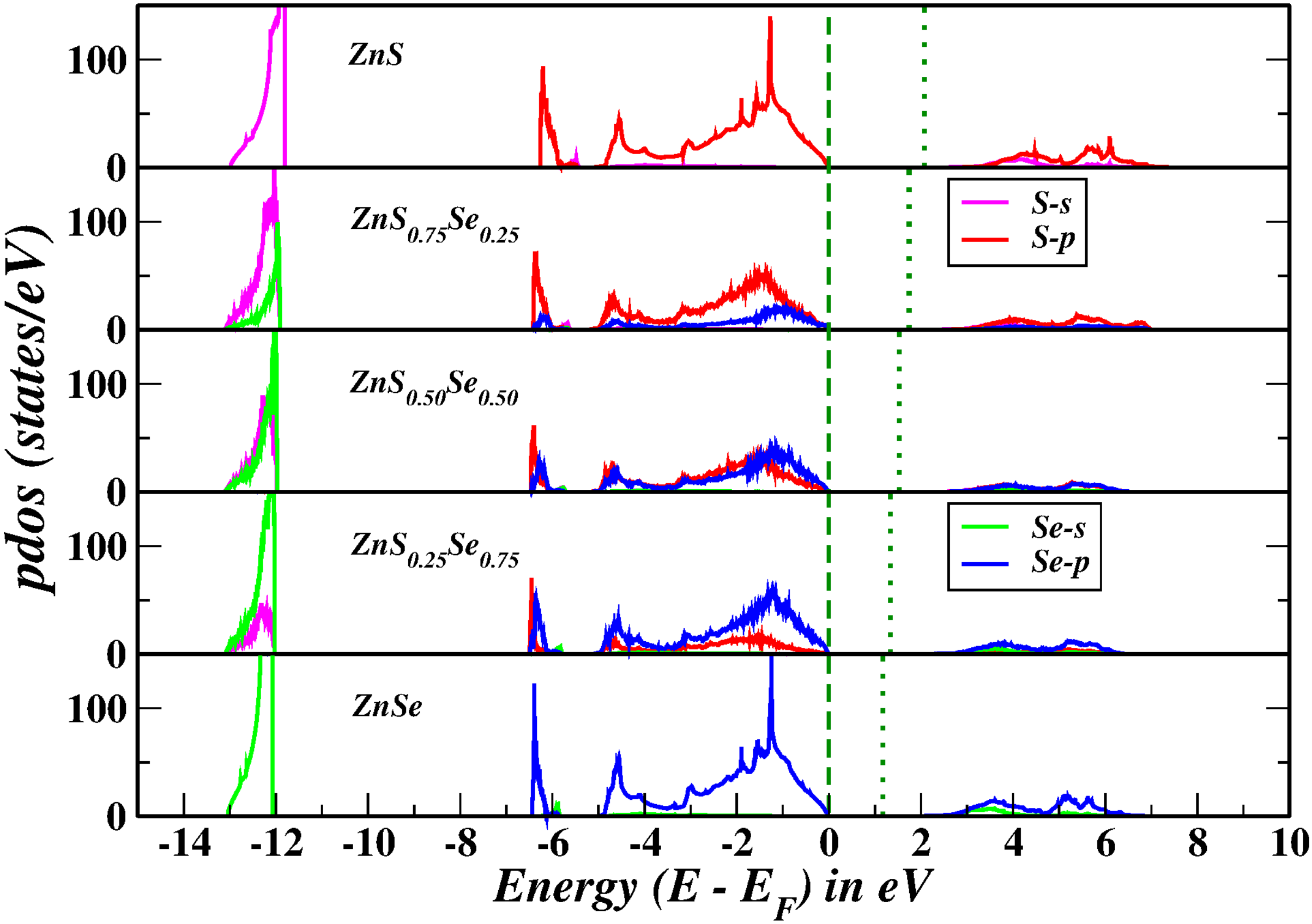}
\caption{Atom projected partial density of states corresponding to the S and Se $s$ and $p$ states in ZnS, ZnSe, and the alloy systems. }
\label{FigC2}
\end{figure}
\FloatBarrier

\end{document}



 \title{Supplemental Material for the manuscript titled\\ Structural and electronic properties of the random alloy ZnSe$_x$S$_{1-x}$ }

\date{\today}
      
\author{S. Sarkar}
\affiliation{Asia Pacific Center for Theoretical Physics, Pohang, 37673, Korea}

\author{O. Eriksson}
\affiliation{Department of Physics and Astronomy, Uppsala University, Box 516, SE-75120, Uppsala, Sweden}
\affiliation{School of Science and Technology, \"{O}rebro University, SE-70182 \"{O}rebro, Sweden}

\author{D.~D. Sarma}
\affiliation{Solid State and Structural Chemistry Unit, Indian Institute of Science, Bengaluru 560012, India}
\affiliation{CSIR-National Institute for Interdisciplinary Science and Technology (CSIR-NIIST), Industrial Estate P.O., Pappanamcode, Thiruvananthapuram 695019, India}

\author{I. Di Marco}
\email{igor.dimarco@apctp.org}
\email{igor.dimarco@physics.uu.se}
\affiliation{Asia Pacific Center for Theoretical Physics, Pohang, 37673, Korea}
\affiliation{Department of Physics and Astronomy, Uppsala University, Box 516, SE-75120, Uppsala, Sweden}
\affiliation{Department of Physics, POSTECH, Pohang, 37673, Korea}

\begin{abstract}  
In this document, we provide additional information about the distributions of the bond lengths and bond angles in the $\text{ZnSe}_{x}\text{S}_{1-x}$ solid solutions, in the wurtzite structure. Histograms for the various distributions in the cubic zinc blende structure are also reported.
\end{abstract}

\maketitle

\setcounter{table}{0}
\renewcommand{\thetable}{S\arabic{table}}

\section{Additional data on the bond lengths}
For an easier comparison, here we report tables illustrating range and average value of the bond lengths in various coordination shells for each composition in the $\text{ZnSe}_{x}\text{S}_{1-x}$ solid solutions, in the wurtzite structure. These tables contain a more compact overview of the data included in the histograms presented in the main text.

\FloatBarrier

\begin{table}[]
\caption{The range and average value of the nearest neighbor Zn-S and Zn-Se bond lengths (\r{A}) for each composition in the $\text{ZnSe}_{x}\text{S}_{1-x}$ solid solutions in the wurtzite structure.}
\label{TableS1}
\begin{tabular}{|l|l|l|l|l|}
\hline
\multirow{2}{*}{Se content (\%)} & \multicolumn{2}{c|}{Zn-S bond (\r{A})} & \multicolumn{2}{c|}{Zn-Se bond (\r{A})} \\ \cline{2-5} 
                  & \multicolumn{1}{c|}{Range}&\multicolumn{1}{c|}{Average}       & \multicolumn{1}{c|}{Range}&\multicolumn{1}{c|}{Average}\\ \hline
     0 \%             & 2.359 - 2.364  & 2.360 & --            & --    \\ \hline
     25 \%            & 2.342 - 2.392  & 2.369 & 2.434 - 2.478 & 2.460 \\ \hline
     50 \%            & 2.352 - 2.407  & 2.377 & 2.448 - 2.499 & 2.470 \\ \hline
     75 \%            & 2.367 - 2.407  & 2.384 & 2.458 - 2.504 & 2.479 \\ \hline
     100 \%           & --             & --    & 2.485 - 2.491 & 2.486 \\ \hline
\end{tabular}
\end{table}

\begin{table}[h]
\caption{ The range and average value of the second neighbor Zn-Zn bond lengths (\r{A}) for all the compositions in the $\text{ZnSe}_{x}\text{S}_{1-x}$ solid solutions  in the wurtzite structure. }
\label{TableS2}
\begin{tabular}{|l|c|c|c|c|c|c|}
\hline
\multirow{2}{*}{Se content (\%)} & \multicolumn{2}{c|}{Zn-Zn bond (\r{A})} & \multicolumn{2}{c|}{\begin{tabular}[c]{@{}l@{}}Zn-Zn bond from  \\ SZn$_4$ units (\r{A})\end{tabular}} & \multicolumn{2}{c|}{\begin{tabular}[c]{@{}l@{}}Zn-Zn bond from \\ SeZn$_4$ units (\r{A})\end{tabular}} \\ \cline{2-7} 
                                 & Range                          & Average                 & Range                                                         & Average                                                 & Range                                                         & Average                                                 \\ \hline
0 \%                             & 3.848 - 3.859                  & 3.854                   & 3.848 - 3.859                                                 & 3.854                                                   & --                                                            & --                                                      \\ \hline
25 \%                            & 3.756 - 4.102                  & 3.906                   & 3.756 - 3.946                                                 & 3.869                                                   & 3.962 – 4.102                                                 & 4.024                                                   \\ \hline
50 \%                            & 3.775 - 4.131                  & 3.957                   & 3.775 - 3.956                                                 & 3.879                                                   & 3.957 – 4.131                                                 & 4.035                                                   \\ \hline
75 \%                            & 3.829 - 4.187                  & 4.009                   & 3.829 - 3.953                                                 & 3.889                                                   & 3.961 – 4.187                                                 & 4.046                                                   \\ \hline
100 \%                           & 4.053 - 4.068                  & 4.060                   & --                                                            & --                                                      & 3.961 – 4.187                                                 & 4.060                                                   \\ \hline
\end{tabular}
\end{table}

\begin{table}[h]
\caption{The range and average value of the second neighbor S-S, Se-Se and S-Se bond lengths (\r{A}) for all the compositions in the $\text{ZnSe}_{x}\text{S}_{1-x}$ solid solutions  in the wurtzite structure. }
\label{TableS3}
\begin{tabular}{|l|c|c|c|c|c|c|}
\hline
\multirow{2}{*}{Se content (\%)} & \multicolumn{2}{l|}{S-S bond (\r{A})} & \multicolumn{2}{l|}{Se-Se bond (\r{A})} & \multicolumn{2}{l|}{S-Se bond (\r{A})} \\ \cline{2-7} 
                                 & Range                         & Average                & Range                          & Average                 & Range                         & Average                 \\ \hline
0 \%                             & 3.848 - 3.859                 & 3.854                  & --                             & --                      & --                            & --                      \\ \hline
25 \%                            & 3.829 - 3.977                 & 3.892                  & 3.854 - 4.012                  & 3.953                   & 3.830 - 3.993                 & 3.915                   \\ \hline
50 \%                            & 3.827 - 4.010                 & 3.929                  & 3.876 - 4.087                  & 3.984                   & 3.864 - 4.037                 & 3.956                   \\ \hline
75 \%                            & 3.927 – 4.017                 & 3.966                  & 3.931 - 4.102                  & 4.022                   & 3.913 - 4.084                 & 3.994                   \\ \hline
100 \%                           & --                            & --                     & 4.053 - 4.068                  & 4.060                   & --                            & --                      \\ \hline
\end{tabular}
\end{table}

\begin{table}[ht]
\caption{The range and average value of the third  neighbor Zn-S and Zn-Se bond lengths for all the compositions in the $\text{ZnSe}_{x}\text{S}_{1-x}$ solid solutions  in the wurtzite structure. }
\label{TableS4}
\begin{tabular}{|c|c|c|c|c|}
\hline
\multirow{2}{*}{Se content (\%)} & \multicolumn{2}{c|}{Zn-S (\r{A})}                                                                                          & \multicolumn{2}{c|}{Zn-Se (\r{A})}                                                                                         \\ \cline{2-5} 
                                 & Range                                                                               & Average                                               & Range                                                                               & Average                                               \\ \hline
0 \%                             & \begin{tabular}[c]{@{}c@{}}Small: 3.948 – 3.948\\ Large: 4.514 – 4.516\end{tabular} & \begin{tabular}[c]{@{}c@{}}3.947\\ 4.515\end{tabular} & --                                                                                  & --                                                    \\ \hline
25 \%                            & \begin{tabular}[c]{@{}c@{}}Small: 3.871 – 4.137\\ Large: 4.440 – 4.736\end{tabular} & \begin{tabular}[c]{@{}c@{}}4.002\\ 4.574\end{tabular} & \begin{tabular}[c]{@{}c@{}}Small: 3.842 – 4.102\\ Large: 4.408 – 4.723\end{tabular} & \begin{tabular}[c]{@{}c@{}}3.995\\ 4.578\end{tabular} \\ \hline
50 \%                            & \begin{tabular}[c]{@{}c@{}}Small: 3.930 – 4.194\\ Large: 4.455 – 4.786\end{tabular} & \begin{tabular}[c]{@{}c@{}}4.061\\ 4.634\end{tabular} & \begin{tabular}[c]{@{}c@{}}Small: 3.860 – 4.196\\ Large: 4.450 – 4.820\end{tabular} & \begin{tabular}[c]{@{}c@{}}4.048\\ 4.637\end{tabular} \\ \hline
75 \%                            & \begin{tabular}[c]{@{}c@{}}Small: 4.003 – 4.280\\ Large: 4.545 – 4.828\end{tabular} & \begin{tabular}[c]{@{}c@{}}4.117\\ 4.693\end{tabular} & \begin{tabular}[c]{@{}c@{}}Small: 3.986 – 4.260\\ Large: 4.517 – 4.866\end{tabular} & \begin{tabular}[c]{@{}c@{}}4.107\\ 4.696\end{tabular} \\ \hline
100 \%                           & --                                                                                  & --                                                    & \begin{tabular}[c]{@{}c@{}}Small: 4.164 – 4.164\\ Large: 4.754 – 4.757\end{tabular} & \begin{tabular}[c]{@{}c@{}}4.164\\ 4.756\end{tabular} \\ \hline
\end{tabular}
\end{table}

\begin{table}[]
\caption{The range and average value of the fourth neighbor Zn-Zn bond lengths (\r{A}) for all the compositions in the $\text{ZnSe}_{x}\text{S}_{1-x}$ solid solutions  in the wurtzite structure.}
\label{TableS5}
\begin{tabular}{|c|cc|}
\hline
\multirow{2}{*}{Se content (\%)} & \multicolumn{2}{c|}{Zn-Zn bond (\r{A})} \\ \cline{2-3} 
 & \multicolumn{1}{c|}{Range} & Average \\ \hline
0 \% & \multicolumn{1}{c|}{5.450 - 5.450} & 5.450 \\ \hline
25 \% & \multicolumn{1}{c|}{5.344 - 5.726} & 5.523 \\ \hline
50 \% & \multicolumn{1}{c|}{5.354 - 5.853} & 5.595 \\ \hline
75 \% & \multicolumn{1}{c|}{5.500 - 5.893} & 5.669 \\ \hline
100 \% & \multicolumn{1}{c|}{5.742 - 5.742} & 5.742 \\ \hline
\end{tabular}
\end{table}

\FloatBarrier

\section{Additional data on the bond angles}
For an easier comparison, here we report some tables illustrating range and average value of the bond angles in various coordination shells for each composition in the $\text{ZnSe}_{x}\text{S}_{1-x}$ solid solutions, in the wurtzite structure. These tables contain a more compact overview of the data included in the histograms presented in the main text.

\FloatBarrier

\begin{table}[h]
\caption{The range and average value of the $X$-Zn-$X$ ($X$ = S or Se) bond angles for all the compositions in the $\text{ZnSe}_{x}\text{S}_{1-x}$ solid solutions in the wurtzite structure. }
\label{TableS6}
\begin{tabular}{|l|c|c|c|c|c|c|}
\hline
\multirow{2}{*}{Se content (\%)} & \multicolumn{2}{c|}{S-Zn-S (Deg)} & \multicolumn{2}{c|}{S-Zn-Se (Deg)} & \multicolumn{2}{c|}{Se-Zn-Se (Deg)} \\ \cline{2-7} 
                                 & Range                         & Average                & Range                          & Average                 & Range                         & Average                 \\ \hline
0 \%                             & 109.32 - 109.62                 & 109.47                  & --                             & --                      & --                            & --                      \\ \hline
25 \%                            & 107.39 - 114.46                 & 110.35                 & 105.25 - 112.58                  & 108.44                   & 103.82 - 109.46                & 107.33                   \\ \hline
50 \%                            & 107.04 - 115.22                 & 111.16                 & 105.54 - 113.57                  & 109.43                  & 103.34 - 112.83                 & 107.74                  \\ \hline
75 \%                            & 110.16 – 113.91                 & 112.20                  & 106.93 - 113.87                  & 110.26                   & 104.90 - 112.28                & 108.60                   \\ \hline
100 \%                           & --                            & --                     & --                            & --                            & 109.26 - 109.68                           & 109.47                   \\ \hline
\end{tabular}
\end{table}

\begin{table}[h]
\caption{The range and average value of the Zn-S-Zn and Zn-Se-Zn bond angles for all the compositions in the $\text{ZnSe}_{x}\text{S}_{1-x}$ solid solutions in the wurtzite structure. }
\label{TableS7}
\begin{tabular}{|l|l|l|l|l|}
\hline
\multirow{2}{*}{Se content (\%)} & \multicolumn{2}{c|}{Zn-S-Zn (Deg)} & \multicolumn{2}{c|}{Zn-Se-Zn (Deg)} \\ \cline{2-5} 
                  & \multicolumn{1}{c|}{Range}&\multicolumn{1}{c|}{Average}       & \multicolumn{1}{c|}{Range}&\multicolumn{1}{c|}{Average}\\ \hline
     0 \%             & 109.32 - 109.62  & 109.47 & --            & --    \\ \hline
     25 \%            & 105.71 - 112.20  & 109.47 & 104.70 - 112.42 & 109.46 \\ \hline
     50 \%            & 105.77 - 113.49  & 109.46 & 105.23 - 113.62 & 109.46 \\ \hline
     75 \%            & 107.15 - 112.93  & 109.47 & 105.76 - 114.10 & 109.46 \\ \hline
     100 \%           & --               & --     & 109.26 - 109.68 & 109.47 \\ \hline
\end{tabular}
\end{table}

\FloatBarrier

\newpage

\section{Histograms for the zinc blende phase}
In the main text, we reported selected data for the zinc blende structure. Here, for sake of completeness, we include all remaining histograms for a proper comparison with the wurtzite structure. They cover data for the bond lengths, bond angles and tetrahedral volumes across various coordination shells for each composition in the $\text{ZnSe}_{x}\text{S}_{1-x}$ solid solutions.

\setcounter{figure}{0}
\renewcommand{\thefigure}{S\arabic{figure}}

\FloatBarrier

\begin{figure}[]
    \includegraphics[scale=0.6]{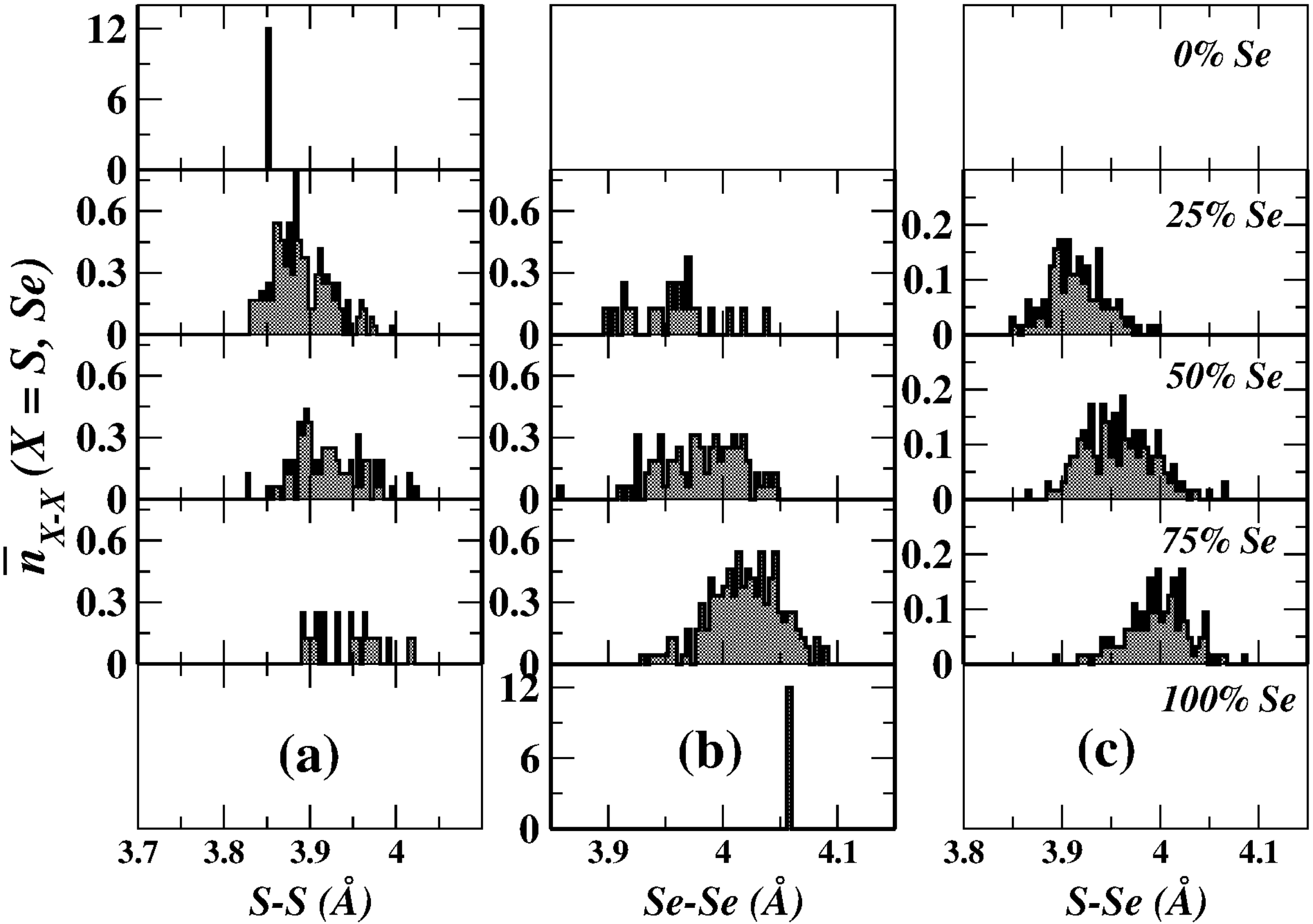}
  \caption{ Distribution of the (a) S-S, (b) Se-Se, and (c) S-Se second neighbor bonds in the cubic phase solid solution alloy for increasing Se concentration (top to bottom), normalized with respect to the number of bonds per S/Se atom (12), weighted by their relative concentration.} 
  \label{FigS1}  
\end{figure}

\begin{figure}[]

    \includegraphics[scale=0.5]{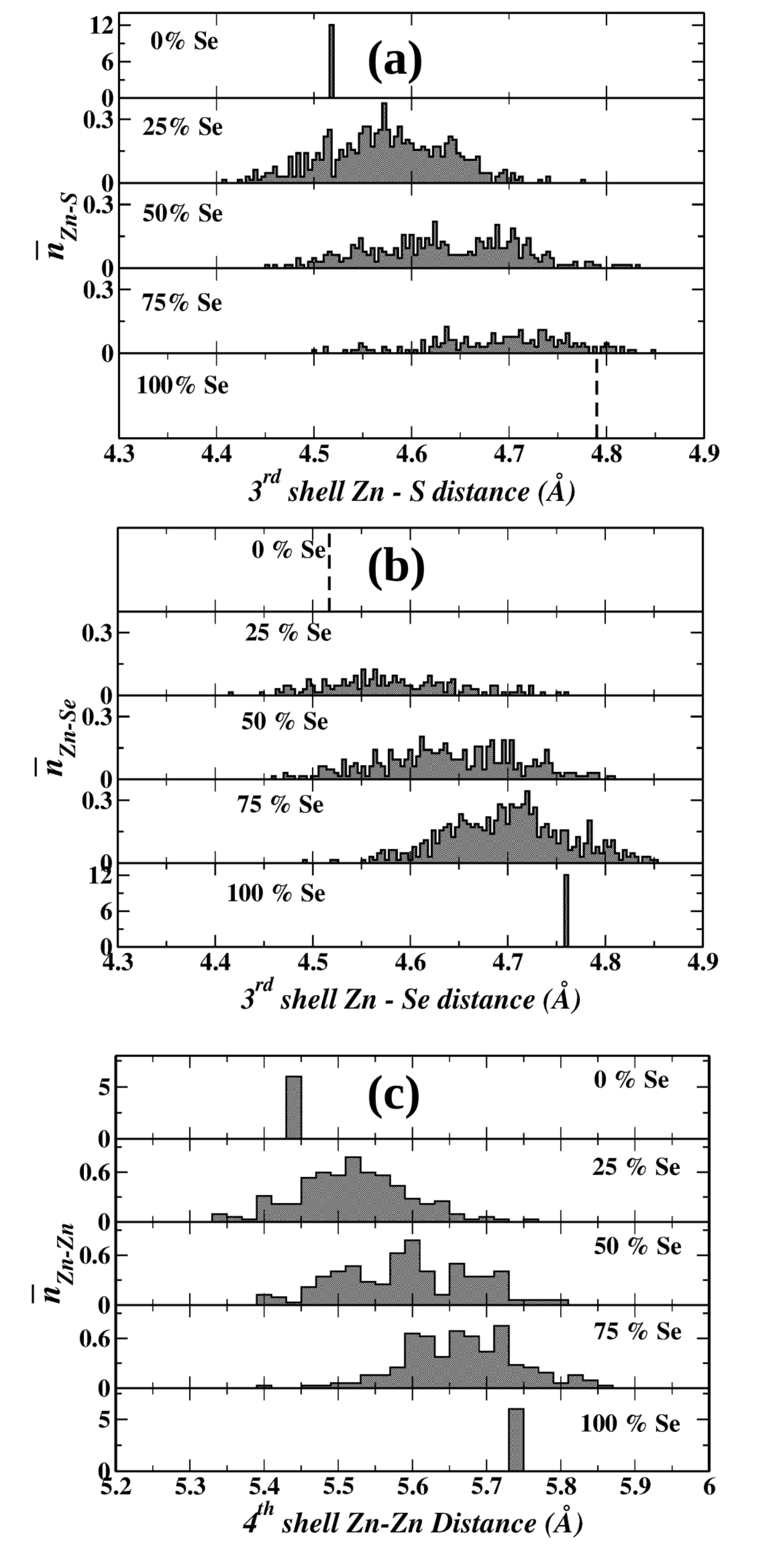}
  \caption{(a) Distributions of the third neighbor Zn-S bonds in the cubic phase of the solid solution alloy for increasing Se concentration (top to bottom), normalized with respect to the number of bonds per Zn atom (12). (b) Distributions of the third neighbor Zn-Se bonds in the cubic phase of the solid solution alloy for increasing Se concentration (top to bottom), normalized with respect to the number of bonds per Zn atom (12). (c) Distributions of the fourth neighbor Zn-Zn bond distances in the cubic phase of the solid solution alloy for increasing Se concentration (top to bottom), normalized with respect to the number of bonds per Zn atom (6).}
  \label{FigS2} 
\end{figure}

 \begin{figure}[ht]
\includegraphics[scale=0.5]{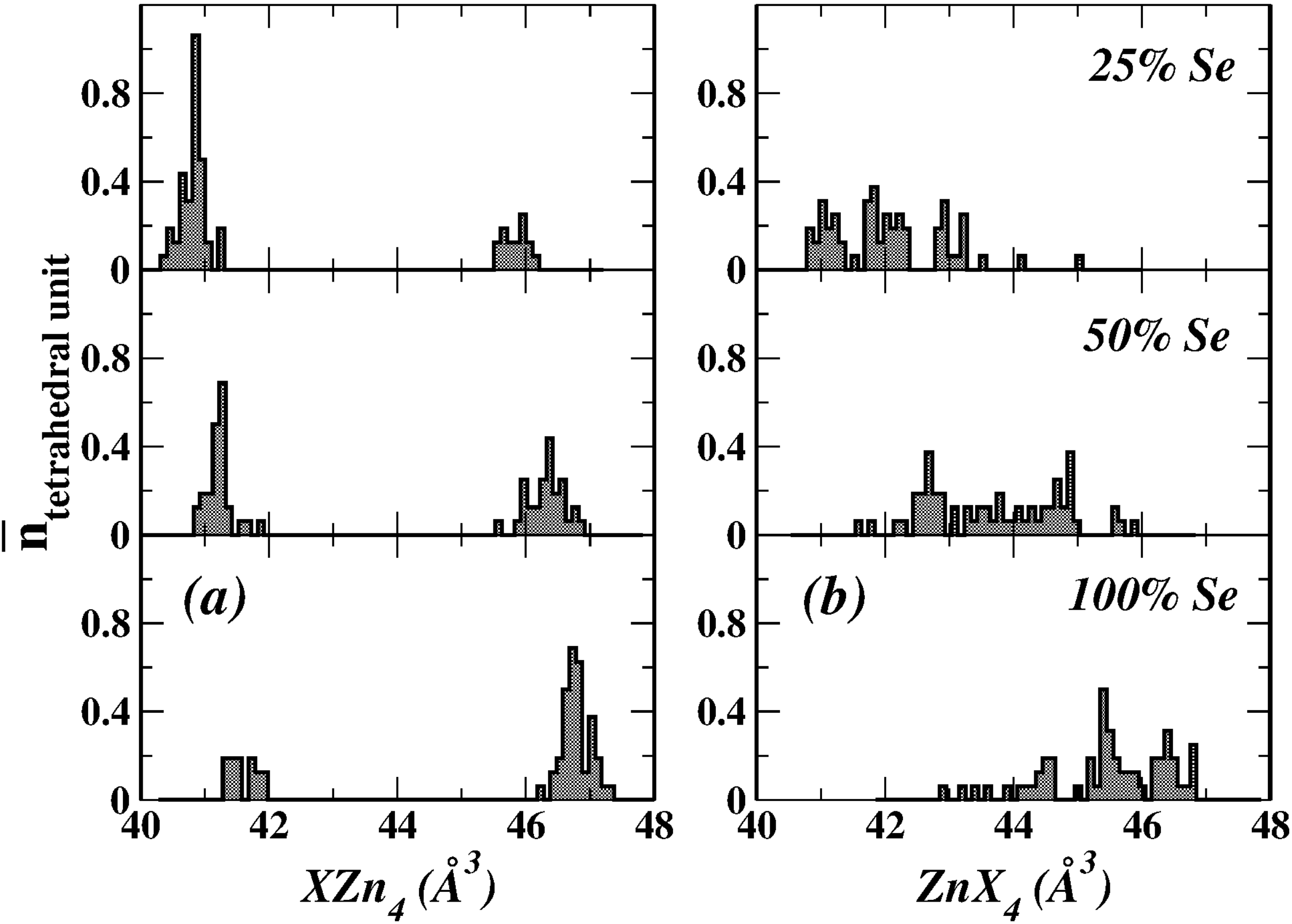}
\caption{Distribution of the calculated volumes associated with the (a) $X$Zn$_4$, and (b) Zn$X_4$ units (X = S or Se) in the cubic phase alloy systems, normalized to 1.}
\label{Fig9}
\end{figure}

\begin{figure}[]

    \includegraphics[scale=0.5]{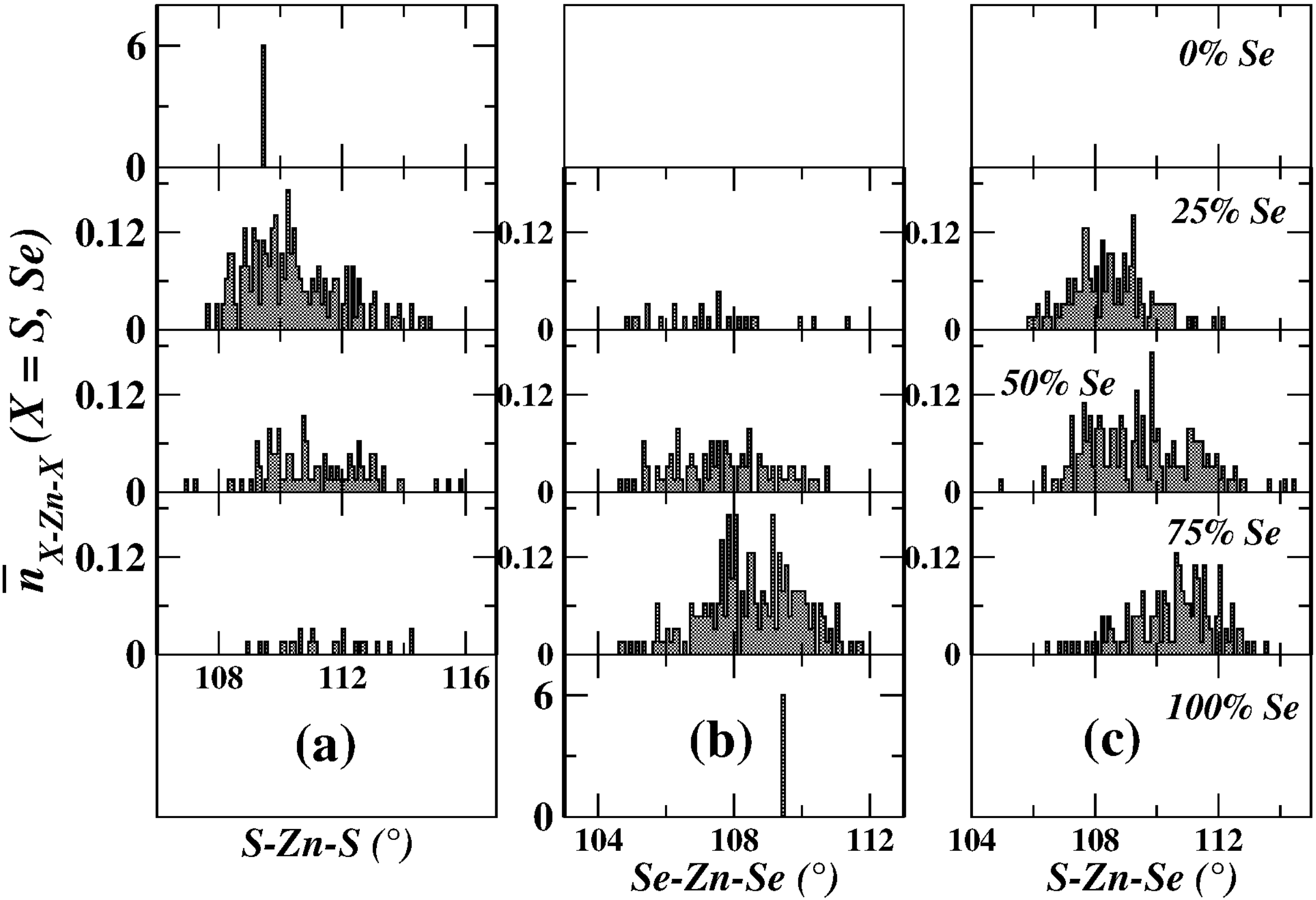}
  \caption{(a) Distributions of the S-Zn-S bond angles in the cubic phase of the solid solution alloy for increasing Se concentration (top to bottom), normalized with respect to the number of bond angles centered on a Zn atom (6) weighted by the S concentration. (b) Distributions of the Se-Zn-Se bond angles in the cubic phase of the solid solution alloy for increasing Se concentration (top to bottom), normalized with respect to the number of bond angles centered on a Zn atom (6) weighted by the Se concentration. (c) Distributions of the S-Zn-Se bond angles in the cubic phase of the solid solution alloy for increasing Se concentration (top to bottom), normalized with respect to the number of bond angles centered on a Zn atom (6) weighted by the pair concentration of S and Se.}
  \label{FigB1}
\end{figure}

\begin{figure}[]

    \includegraphics[scale=0.5]{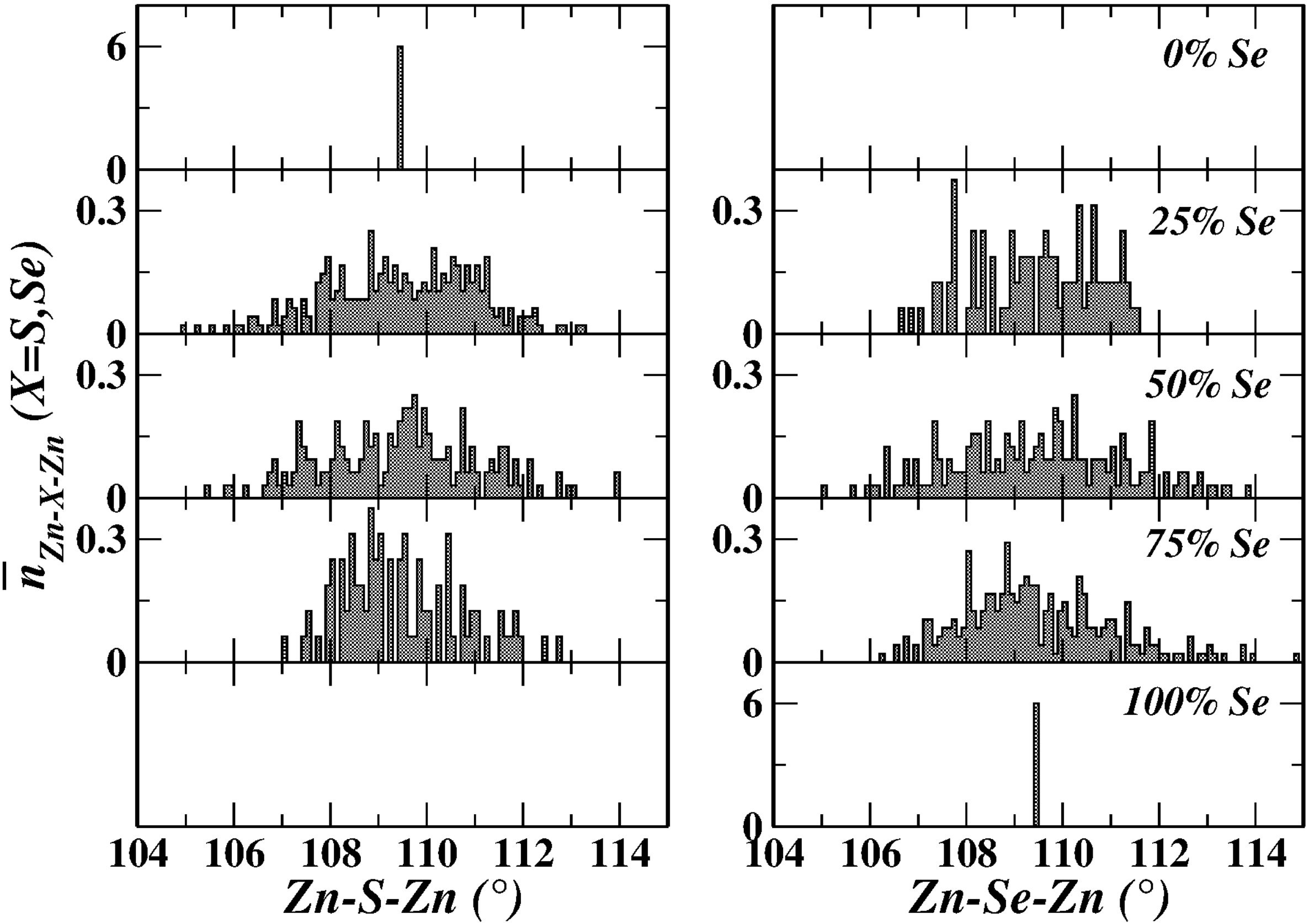}
  \caption{(a) Distributions of the Zn-S-Zn bond angles in the cubic phase of the solid solution alloy for increasing Se concentration (top to bottom), normalized with respect to the number of bond angles centered on a S atom (6). (b) Distributions of the Zn-Se-Zn bond angles in the cubic phase of the solid solution alloy for increasing Se concentration (top to bottom), normalized with respect to the number of bond angles centered on a Se atom (6).}
  \label{FigB2}
\end{figure}

\FloatBarrier